\documentclass[
 reprint,
 amsmath,amssymb,
 aps,
 superscriptaddress,
 nofootinbib,
]{revtex4-2}

 \usepackage[skins,theorems]{tcolorbox}
 \usepackage{fontawesome5}
 \newtcbox{\ghbadge}{on line, arc=3pt, colback=gray!12, colframe=gray!40, 
  boxrule=0.5pt, boxsep=0pt, left=4pt, right=4pt, top=2pt, bottom=2pt}
\tcbset{highlight math style={enhanced,
  colframe=red,colback=white,arc=0pt,boxrule=1pt}}
  \usepackage[bookmarksopen, bookmarksnumbered, bookmarksopenlevel=2]{hyperref}
  \usepackage{tikz}
  \usepackage{tikz-3dplot}
  \usepackage{float}
 \usetikzlibrary{calc}
 \usetikzlibrary{decorations}
 \usepackage[UKenglish]{babel}
 \usepackage[toc,page]{appendix}
 \usepackage{amsmath}
 \usepackage{amssymb}
 \usepackage{graphicx}
 \usepackage{hhline}
 \usepackage[bf]{caption}
\usepackage[table]{xcolor} 
\usepackage{colortbl}       
\usepackage[vcentermath]{youngtab}
\usepackage{geometry}
\usepackage{slashed}
\usepackage{color}
\usepackage{stackrel}
\usepackage{tikz-cd} 
\usepackage{mathtools}
\usepackage{cancel} 
\usepackage{multirow}
\usepackage[margin=0pt,font=small,labelfont=normalfont,skip=22pt]{subcaption}
\usepackage{empheq}
\usepackage{color,soul}
\usepackage{physics}
\usepackage{amsmath}
\usepackage{tikz}
\usepackage{mathdots}
\usepackage{yhmath}
\usepackage{cancel}
\usepackage{color}
\usepackage{siunitx}
\usepackage{array}
\usepackage{float}
\usepackage{multirow}
\usepackage{amssymb}
\usepackage{gensymb}
\usepackage{tabularx}
\usepackage{extarrows}
\usepackage{booktabs}
\usetikzlibrary{fadings}
\usetikzlibrary{patterns}
\usetikzlibrary{shadows.blur}
\usetikzlibrary{shapes}
\usepackage{cleveref}

\newcommand{\bqa}{\begin{eqnarray}}
\newcommand{\eqa}{\end{eqnarray}}

\definecolor{darkgreen}{RGB}{50,150,0}

\newenvironment{eqn*}{\begin{equation*}\begin{aligned}}{\end{aligned}\end{equation*}\noindent}
\hypersetup{
    pdftitle={},
    pdfauthor={},
    pdfsubject={}
}
\makeatletter

\newcommand{\be}{\begin{equation}}
\newcommand{\ee}{\end{equation}}
\newcommand{\beq}{\begin{equation}}
\newcommand{\eeq}{\end{equation}}
\newcommand{\ba}{\begin{aligned}}
\newcommand{\ea}{\end{aligned}}

\newcommand{\bea}{\begin{eqnarray}}
\newcommand{\eea}{\end{eqnarray}}

\newcommand{\cL}{\mathcal{L}}
\newcommand{\cK}{\mathcal{K}}

\newcommand{\cG}{\mathcal{G}}
\newcommand{\cA}{\mathcal{A}}
\newcommand{\cH}{\mathcal{H}}

\newlength{\myfootnotewidth}
\newcommand\bi{\begin{itemize}}
\newcommand\ei{\end{itemize}}

\def\unit{{1\kern-.65ex {\rm l}}}
\def\1{{1\kern-.65ex {\rm l}}}

\def\bk{{\mathbf{k}}}
\def\bq{{\mathbf{q}}}
\def\bp{{\mathbf{p}}}

\def\dd{{\mathrm{d}}}
\def\pl{{\mathrm{pl}}}

\newcount\hour \newcount\minute
\hour=\time \divide \hour by 60
\minute=\time
\def\now{%
\ifnum \hour<13
  \ifnum \hour=0 \advance \hour by 12 \number\hour:\else \number\hour:\fi%
     \ifnum \minute<10 0\fi%
     \number\minute%
\ A.M.%
\else \advance \hour by -12 \number\hour:%
  \ifnum \minute<10 0\fi%
  \number\minute%
  \ P.M.%
\fi%
}

\def\p@subsection{}
\def\p@subsubsection{}

\makeatother

\begin{document}

\title{The Full-Shape Analysis of the Fading Dark Sector Model}

\author{Georges Obied}
\affiliation{Department of Physics and Astronomy,
University of North Carolina at Chapel Hill,
Chapel Hill, NC 27599, USA}

\author{Cumrun Vafa}
\affiliation{Jefferson Physical Laboratory, Harvard University, Cambridge, MA 02138, USA}

\author{David H. Wu}
\affiliation{Jefferson Physical Laboratory, Harvard University, Cambridge, MA 02138, USA}

\date{\today}

\begin{abstract}
We analyze the Fading Dark Sector (FDS) model, which is motivated by string theory and in which the fading of the dark sector is driven by a scalar field. The model is specified by two $\mathcal{O}(1)$ numbers $(c,c')$ that parametrize the logarithmic derivatives of the dark energy and dark matter densities with respect to the field in Planck units. In previous work, we showed that this model provides a good fit to the DESI BAO, CMB, and SN datasets. Here, we extend that analysis to include the DESI DR1 full-shape galaxy power spectrum, which adds information about the time evolution of cosmic structure and provides an independent consistency check of the FDS model. We find that, when the DESI DR1 full-shape data are combined with CMB+DESI DR2 BAO+DES-Dovekie, the FDS model outperforms the Chevallier--Polarski--Linder (CPL) $w_0$--$w_a$ parameterization by $\Delta \chi^2_{\rm MAP}=-4$ at equal parameter count and is preferred over $\Lambda$CDM at $4.1\sigma$. The inferred values of $(c,c')$ with the full-shape data are unchanged from our previous analysis, so this improvement indicates that FDS makes a prediction for the growth of structure that captures the observed growth history better than the $w_0$--$w_a$ parameterization.
\end{abstract}

\maketitle

\section{Introduction}

Recent measurements of the baryon acoustic oscillation (BAO) scale by DESI are in $2-3\sigma$ tension with predictions of the standard $\Lambda$CDM model inferred from observations of the cosmic microwave background (CMB)~\cite{DESI:2024mwx,DESI:2025zgx}. This tension is commonly described as a preference by DESI for a smaller matter fraction $\Omega_m$ than the one obtained from CMB measurements. Additionally, supernova samples~\cite{Brout:2022vxf,Rubin:2023jdq,DES:2024jxu,DES:2025sig} have their own different preference for a value of $\Omega_m$ that is larger than the one preferred by the CMB. This means that excluding any single dataset still leaves a tension within $\Lambda$CDM. Various modifications to the standard cosmological model have been proposed to explain these discrepancies (see for example~\cite{DESI:2025fii} and references therein). These include phenomenological and astrophysical approaches such as the possibility of having a larger optical depth or replacing the cosmological constant with a dark energy fluid, parametrized by its equation of state (such as $w_0w_a$ or more general variations such as in~\cite{Shlivko:2025fgv}). Other attempts include modifications to the physics such as spatial curvature, non-minimal couplings, modified gravity, modified neutrino sector or interacting dark sectors~\cite{Agrawal:2019dlm,Khoury:2025txd,Bagherian:2026jxz}. The Fading Dark Sector model (FDS) studied here, which was motivated from string theory (rather than DESI data) and proposed in 2019 \cite{Agrawal:2019dlm}, belongs to the latter category. 

Several of these models are degenerate at the level of the background cosmology, which makes them indistinguishable using distance measurements alone, such as the BAO scale. For example, distance measurements can be used to infer an effective $w(z)$ for a dark energy fluid but several underlying physical theories can give rise to the same expansion history and hence the same effective $w(z)$ despite having important differences in their microphysics (see for example~\cite{Ishak:2024jhs}). The full shape (FS) of the galaxy power spectrum contains additional information that can break this degeneracy between models~\cite{DESI:2024hhd}. This information arises from redshift space distortions (which give sensitivity to the growth rate), the Alcock–Paczynski effect (which gives geometric information on the expansion) and the broadband shape of the power spectrum (which can determine the scale of equality). Altogether, full-shape (FS) analyses can discriminate between models that look equivalent at the background level.

In physical models that arise from a concrete modification of the underlying physics, the growth rate of structure, along with the other FS observables mentioned above, are predictions of the model. This is because the same parameters that modify the background expansion also fix the evolution of perturbations. This is unlike phenomenological parameterizations where the behavior of perturbations is an independent assumption that is not tied to $w(z)$. In the typical $w_0$--$w_a$ parameterization, for instance, one assumes that the dark energy fluid is smooth and gravity is general relativity and only after making these assumptions can one calculate modifications to the growth rate of structure. A model that is defined microscopically does not have room for these additional assumptions. FS data can therefore be used as a test which can falsify or disfavor some models relative to others. In particular, there could be models, where the background evolution alleviates the tension between geometric observables, which fail the FS test, e.g., by producing too much structure. In this paper, we perform this test on the fading dark sector (FDS) model.

The FDS model fits well with the cosmology of the Dark Dimension Scenario~\cite{Montero:2022prj} which is motivated by string-theoretic considerations and leads to a unification of the dark sector (see \cite{Vafa:2026ccm} for a review). The FDS model, which we discuss in section~\ref{sec:lss}, is a two parameter extension of $\Lambda$CDM.  Relative to $\Lambda$CDM, it includes an additional scalar field $\phi$ which controls both the dark energy and the dark matter mass. Locally in field space the potential is approximated by an exponential $V\sim V_0\, e^{-c\phi}$ where we are measuring $\phi$ in Planck units. The mass of the dark matter is likewise taken to be locally $m\sim m_0\, e^{-c'\phi}$. The two exponents $(c,c')$ which denote the logarithmic slopes of dark energy and dark matter with respect to $\phi$, are the additional parameters of the model and are inputs from microscopic physics that determine the cosmological evolution.  The most non-trivial aspect of the FDS model, in addition to the fact that it couples the dark energy and dark matter through $\phi$, is that based on expectations from string theory the parameters $(c,c')$ should take $\mathcal{O}(1)$ values.  This feature of the FDS model distinguishes it from a generic interacting dark matter/dark energy model. For example, with $c \sim \mathcal{O}(1)$, one expects $|\nabla V|\sim 10^{-122\pm 1}$ in Planck units with no tuning beyond that of the vacuum energy itself. Indeed fitting FDS to recent cosmological data, such as the DESI DR2 BAO and DES-Dovekie supernova measurements, leads to $O(1)$ parameters for both $(c,c')\sim (-0.8\pm 0.2,0.05\pm 0.01)$~\cite{Bedroya:2025fwh} as predicted in~\cite{Agrawal:2019dlm}.

The evolution of $\phi$ is dictated by the potential during dark energy domination and $\phi$ rolls in the direction that reduces the dark energy density. On the other hand, during matter domination, the evolution of $\phi$ is dictated by the dark matter density and $\phi$ rolls during that era in the direction that decreases the dark matter density. The evolution of $\phi$ therefore naturally causes the dominant dark component, be it the dark matter or dark energy, to decrease and gives rise to the fading dark sector (FDS). 

Our analysis shows that the inclusion of FS data leaves the inferred mean values of $(c,c')$ unchanged from those preferred by geometric distance measures (such as DESI DR2 BAO) and it also tightens the constraints on $(c,c')$ in a way consistent with our previous findings~\cite{Bedroya:2025fwh}. Importantly, not only is the fit not degraded by adding information from the FS power spectrum, but the significance of the model {\it improves} relative to both $\Lambda$CDM as well as the phenomenological CPL $w_0-w_a$ parametrization. We find that, when the DESI DR1 full-shape data are combined with CMB+DESI DR2 BAO+DES-Dovekie, the FDS model outperforms the CPL $w_0$--$w_a$ parameterization by $\Delta \chi^2_{\rm MAP}=-4$ at equal parameter count and is preferred over $\Lambda$CDM at $4.1\sigma$!  This pattern persists even without any SN datasets, i.e., only with DESI and CMB data, which yields a $3.4\sigma$ significance over $\Lambda$CDM and over CPL by $\Delta\chi^2_{\rm MAP}\approx -3.5$.

The rest of the paper is organized as follows.  In section~\ref{sec:lss}, we review the cosmology of the FDS model and the structure of the one-loop power spectrum. In section~\ref{sec:data}, we introduce the data used in this paper.  In section~\ref{sec:results}, we present results from the Markov Chain Monte Carlo analysis of the FDS model and the physics of the best-fit model. We also analyze the growth of structure in the FDS model. In section~\ref{sec:discussion}, we end with concluding remarks.  We relegate the derivation of the equations of motion governing the one-loop contribution to the multipole moments of the galaxy power spectrum to appendix~\ref{app:eft_derivation} and discuss the mapping of parameters to the DESI FS data in appendix~\ref{sec:DESI multipoles}.

\section{Cosmological dynamics}
\label{sec:lss}

In the fading dark sector, the dark energy and dark matter are coupled through a scalar field
\begin{equation}
    V_{\rm DE}=V_0\, e^{-c\phi}\,,\qquad m_{\rm DM}=m_0\, e^{-c'\phi}\,,
\end{equation}
where $\phi$ is a dimensionless scalar field. As we have discussed in the introduction, the scalar field evolves with the effective potential $V_{\rm eff}=V_0e^{-c\phi}+a^{-3}\rho^{DM}_{\rm 0}e^{-c'\phi}$ which leads to fading of dark matter (in matter dominated era) and dark energy (in the dark energy dominated era). The scalar in the FDS is light, so that it is able to roll on cosmological scales, and its coupling to dark matter means that it can mediate a long-range fifth force in the dark sector. On subhorizon scales the scalar perturbation can be treated in the
quasi-static approximation. Its effect is an additional long-range fifth force (with strength $2c'^2$) to gravitational force acting only on the dark matter:
\begin{equation}
\label{eq:main_Geff}
    G_{\chi\chi}^{\rm eff}(k,a)=G_N\left[1+\Gamma(k,a)\right]\,,
\end{equation}
where $\chi$ denotes the coupled dark matter component and 
\begin{equation}
\label{eq:main_Gamma}
    \Gamma(k,a)=\frac{2c'^2k^2}{k^2+a^2m_{\rm eff}^2(a)}\,,
\end{equation}
with $m_{\rm eff}$ the effective mass of the scalar perturbation. Thus, on short scales, $k\gg am_{\rm eff}$, the force between dark matter particles is enhanced by a factor $1+2c'^2$ as was calculated in \cite{Bedroya:2025fwh}. The $c'$ controlling the strength of the fifth force is also the same parameter that sets the fading of dark matter in the matter dominated era. 
This is an example the lack of freedom in physical models to modify cosmological dynamics separately from the microphysics. 

The coupling also changes the momentum dilution of the dark matter fluid.  It is useful to characterize this effect by the dimensionless quantity
\begin{equation}
\label{eq:main_u}
    u(a)\equiv-c'\frac{\dd\phi}{\dd\ln a}\,.
\end{equation}
The two modifications have physically distinct origins: $\Gamma$
describes the additional scalar force originating from the fluctuation of the scalar field, while $u$ describes the
modified drag associated with the evolving dark matter mass. The
potential parameter $c$ affects structure formation indirectly
through the background evolution of $\phi$, $H$, and the matter
fractions, and therefore also through $u$.

The background and linear perturbation equations for the FDS model have been derived and implemented previously in~\cite{Bedroya:2025fwh}. Here, to probe the physics of the FDS model in the mildly non-linear scale, we adapt the framework of the effective field theory of large-scale structure (EFTofLSS) for the FDS model.
The EFTofLSS for the FDS model is considerably more involved compared to e.g., $\Lambda$CDM. In particular, we cannot rely on the derived Einstein--de Sitter kernels of the EFTofLSS for $\Lambda$CDM due to 1) the baryons and fading dark matter have different equations of motion in the regime of interest and 2) the additional terms arising from the fifth force enhancement $1+\Gamma(k,a)$ and the scalar induced friction $c'\phi_{,\ln a}$ appearing in the modified Euler equations for perturbations of dark matter are time dependent. Hence, we compute the FDS kernels and evolve the system with the exact time dependence. Importantly, for the EFTofLSS for the FDS model, we do not assume that the density and velocity perturbations of the baryon and fading dark matter fluid can be factorized into time-independent kernels, where the time-dependence can be entirely absorbed into the growth factors. Hence, the full system of kernels, including the recursion relation defining higher-order corrections, are all time dependent.\footnote{For the perturbative treatment when only a weak fifth force is added to the dark sector of $\Lambda$CDM, see e.g.,~\cite{Bottaro:2023wkd}. In particular, they have relied on a perturbative correction to the Einstein-de Sitter kernels.}

\subsection{Galaxy power spectrum}

In redshift space, the galaxy power spectrum can be organized as
\begin{multline}
\label{eq:main_Pgs}
    P_g^s(k,\mu)=P_{\rm tree}(k,\mu)+P_{\rm 1\mbox{-}loop}(k,\mu)\\
    +P_{\rm ct}(k,\mu)+P_{\rm stoch}(k,\mu)\,,
\end{multline}
where $\mu\equiv\hat{\boldsymbol{k}}\cdot\hat{\boldsymbol{z}}$ is the cosine of the angle between the Fourier mode $\boldsymbol{k}$ and the line of sight $\hat{\boldsymbol{z}}$. The first term contains the linear density and redshift-space-distortion contribution, $P_{\rm 1\mbox{-}loop}$ contains the standard $P_{22}$ and $P_{13}$ one-loop terms evaluated with our model-specific kernels, $P_{\rm ct}$ contains the EFT counterterms that absorb unresolved short-distance dynamics, and $P_{\rm stoch}$ describes stochastic contributions associated with the discrete galaxy tracers.
The observable multipole moments of the redshift space galaxy power spectrum can then be deduced from $P_g^s(k,\mu)$ as
\begin{equation}
\label{eq:main_multipoles}
    P_\ell(k)=\frac{2\ell+1}{2}\int_{-1}^{1}d\mu\,{\cal L}_\ell(\mu)P_g^s(k,\mu)\,,
\end{equation}
where ${\cal L}_\ell$ is the Legendre polynomial of order $\ell$. We compute the monopole, quadrupole, and hexadecapole, corresponding to $\ell=0,2,4$. 

The detailed derivation of $P_g^s(k,\mu)$ for the FDS model is given in appendix~\ref{app:eft_derivation}. Here, we outline this procedure for computing $P_g^s(k,\mu)$ and consequently $P_l(k)$. We start with deriving the higher-order corrections to the continuity and Euler equations for baryons and the fading dark matter, which are given by \cref{eq:pert fourier 1,eq:pert fourier 2,eq:pert fourier 3}. To solve these equations, we adapt the kernel expansion for the density and velocity perturbation for baryons and dark matter given in \cref{eq:kernel Fs,eq:kernel Gs}. Then, we derive the exact-time dependent recursion relation for the kernels in~\cref{eq:recursion}. With the higher-order corrections computed for the matter power spectrum, we use the two-fluid Eulerian basis for the galaxy bias shown in~\cref{eq:galaxy pert} and use its relation to total and relative matter perturbations (along with higher-order corrections) to derive the relevant kernels appearing in the galaxy density and velocity expansions. Upon mapping the power spectrum to redshift space via~\cref{eq:redshift realspace} to account for the redshift-space distortions, adding the appropriate counter-terms~\eqref{eq:counterterm} and stochastic corrections~\eqref{eq:stochastic}, and lastly performing an IR resummation, we arrive at the final form of the one-loop corrected galaxy power spectrum~\cref{eq:final power spectrum}. As a last step to make contact with galaxy survey measurements, we perform an Alcock--Paczynski mapping given by~\cref{eq:fiducial mapping} which brings galaxy power spectrum into its appropriate form for \cref{eq:main_multipoles}.

\section{Data}
\label{sec:data}

We have implemented the fading dark sector in \verb+CLASS-PT+~\cite{Chudaykin:2020aoj} and we use \verb+Cobaya+~\cite{Torrado:2020dgo} to run the Markov Chain Monte Carlo (MCMC) simulations.

\subsection{DESI full-shape likelihood}

We test our EFTofLSS analysis against the public DESI DR1 full-shape likelihood~\cite{DESI:2024hhd}.\footnote{\href{https://github.com/cosmodesi/desi-kp-cosmological-likelihoods}{https://github.com/cosmodesi/desi-kp-cosmological-likelihoods}} The measured data vectors, covariance
matrices, effective redshifts, scale cuts, and survey window
functions are kept identical to those of the DESI analysis; only the theoretical galaxy power spectrum is replaced by the coupled dark sector description described above. The DESI DR1 galaxy full-shape measurements are carried out for all the BGS, LRG, ELG, and QSO samples. To specify the power spectrum only, we use the option \verb+spectrum-pole-rotated+ to separate the DR1 full-shape likelihood from the DR1 BAO likelihood.

For each galaxy sample, the three nonlinear bias parameters sampled explicitly are
\begin{equation}
\label{eq:M1_bias_main}
    (1+b_1^L)\sigma_8(z)\,,\qquad b_2^L\sigma_8^2(z)\,,\qquad b_s^L\sigma_8^2(z)\,,
\end{equation}
where $b_1^L$ is the Lagrangian linear bias while $b_2^L$ and $b_s^L$ are the quadratic local and tidal Lagrangian bias parameters, respectively, and $\sigma_8(z)$ is the rms amplitude of the linear non-neutrino matter fluctuations in spheres of radius $8\,h^{-1}{\rm Mpc}$ at redshift $z$.

The remaining EFT directions are chosen such that they enter the likelihood linearly.  In particular, adopting the DESI notation, we analytically marginalize over two counterterm amplitudes, $\alpha_0$ and $\alpha_2$, and two stochastic amplitudes, ${\rm SN}_0$ and ${\rm SN}_2$.  Their priors, together with those of the sampled bias parameters, are summarized in Table~\ref{tab:nuisance_priors}.  The nuisance parameters are independent between the six DESI full-shape tracer/redshift samples.

\subsection{Additional datasets}
\label{sec:additional data}

In addition to the DESI DR1 galaxy full-shape likelihoods, we also include CMB temperature, polarization, and lensing measurements, DESI DR1 and DR2 BAO measurements, and the DES-Dovekie Type-Ia supernova compilation.

For the CMB, we combine the Planck low-$\ell$ TT and EE likelihoods~\cite{Planck:2019nip} with the Planck PR4/NPIPE CamSpec high-$\ell$ TT, TE, and EE spectra~\cite{Planck:2020olo,Rosenberg:2022sdy}. We additionally include CMB lensing information from the joint ACT DR6~\cite{ACT:2023kun,ACT:2023dou} and Planck lensing likelihood~\cite{Carron:2022eyg}. This choice is consistent with~\cite{DESI:2025zgx}.

When BAO information is included together with full shape, we use the DESI DR1 FS~\cite{DESI:2024jxi} and DESI DR2 BAO~\cite{DESI:2025zgx} likelihood, which we refer to as DESI1.5. In joint analyses of DESI DR1 full-shape galaxy power spectrum and DESI DR2 BAO, \cite{Chudaykin:2025aux,Forero-Sanchez:2026bff} have found that the cross correlation between the two datasets have little impact on the determination of the cosmological parameters. Hence, for our present simulations, we assume the independence of full-shape and BAO as in~\cite{Chudaykin:2025aux} by imposing the following likelihoods
\begin{equation}
    \ln \cL_{\rm DESI1.5}=\ln \cL_{\text{DR1 FS}}+\ln\cL_{\text{DR2 BAO}}\,.
\end{equation}

Lastly, we also make use of the Type Ia supernova (SN) luminosity distances. In the present paper, we use the recently recalibrated DES-Dovekie sample~\cite{DES:2025sig}. Motivated by our previous findings for the FDS model~\cite{Bedroya:2025fwh}, we expect little deviation for the MCMC and best-fit results among the choice of recalibrated supernova datasets (e.g., recalibrated Pantheon+ or Union3.1~\cite{Hoyt:2026fve,Rubin:2026qdt}).

\section{Results}
\label{sec:results}

\begin{figure*}[!tp]
    \centering
    \begin{subfigure}{.49\textwidth}
    \includegraphics[width=\linewidth]{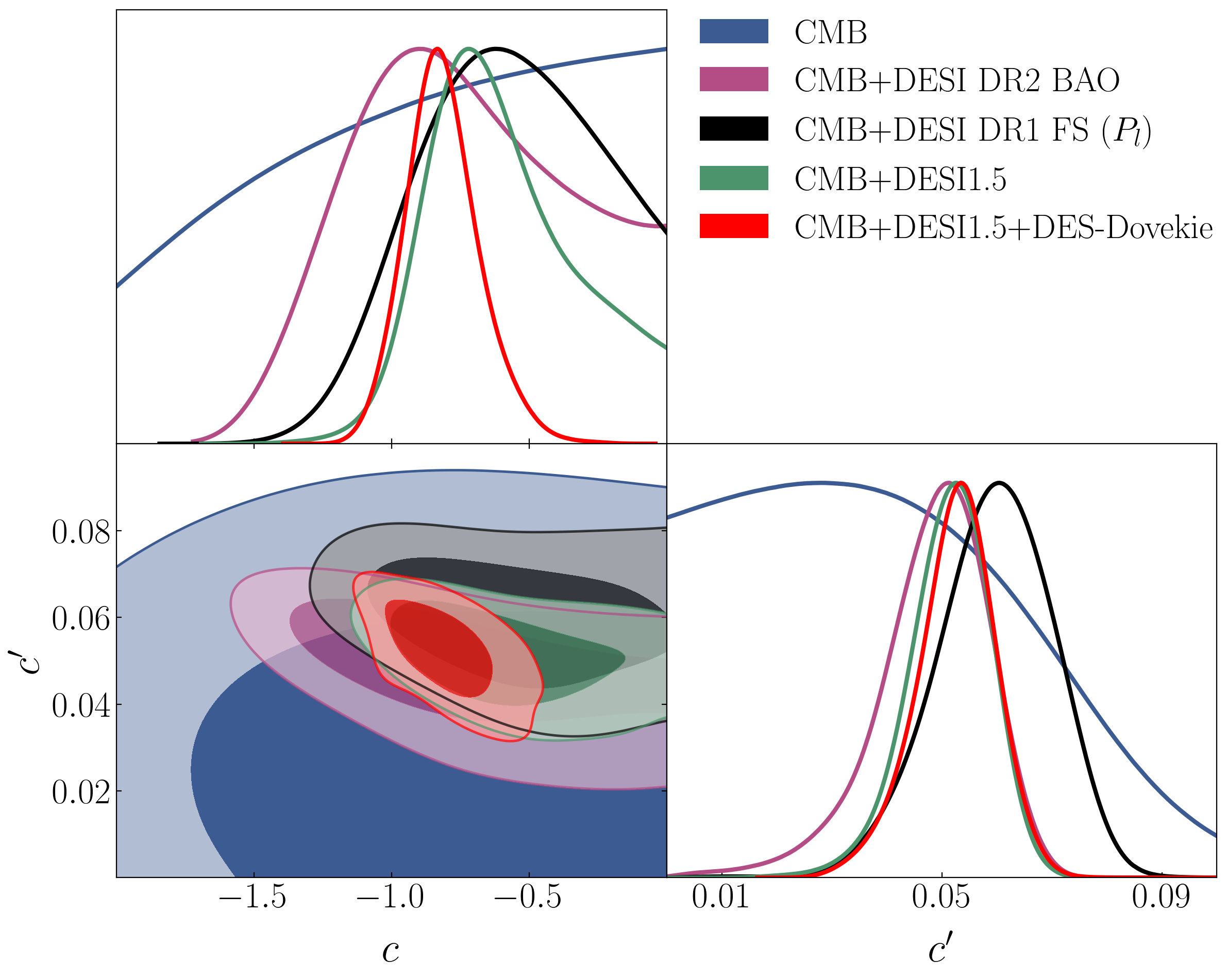}
    \end{subfigure}
    \hfill
    \begin{subfigure}{.49\textwidth}
    \includegraphics[width=\linewidth]{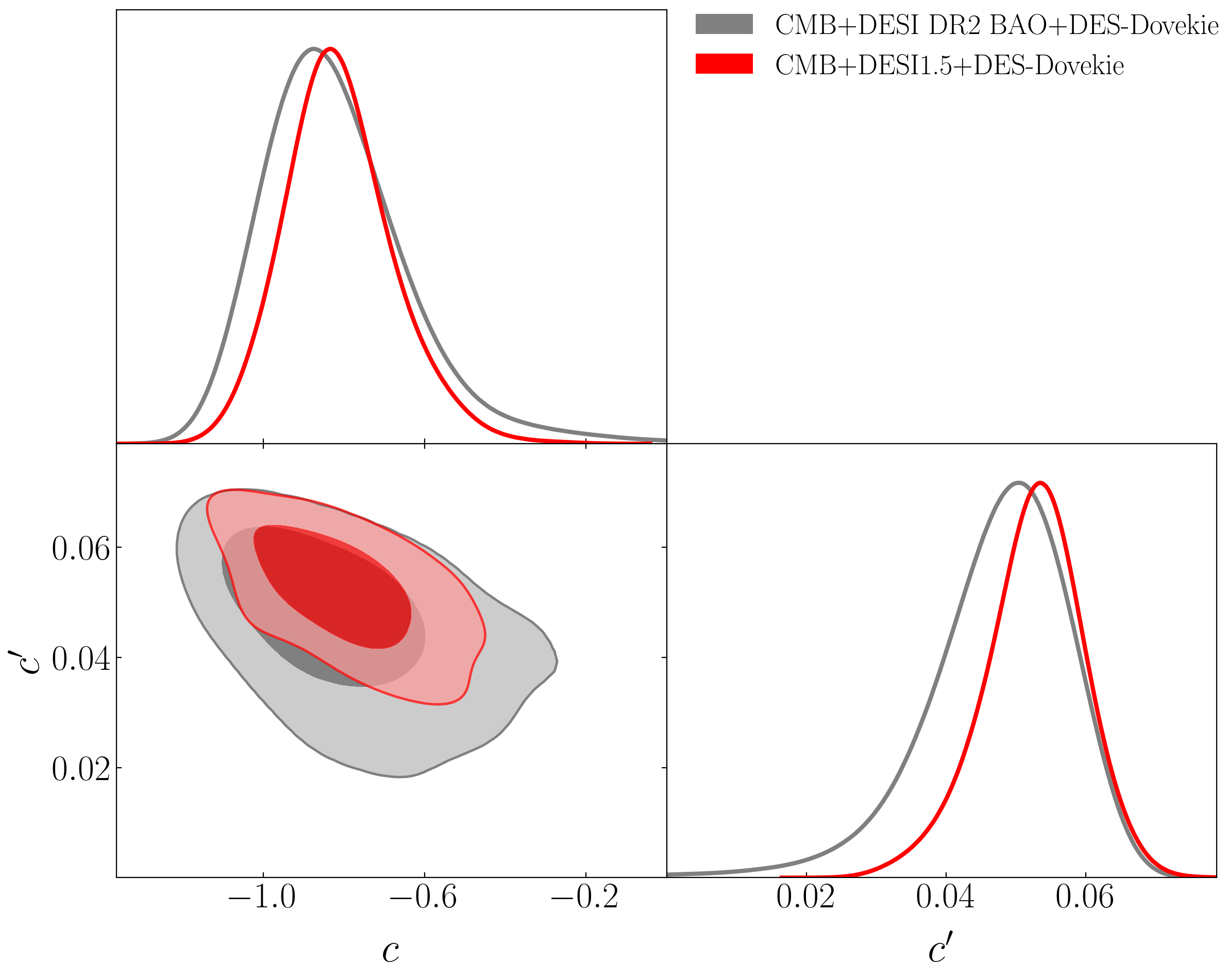}
    \end{subfigure}
    \caption{Contours showing the 1- and 2-$\sigma$ constraints on c and c' for the FDS model across various combination of datasets.}
    \label{fig:ccp}
\end{figure*}

\begin{table*}[!t]
\centering
\begin{tabular}{c|c|>{\hspace{.2cm}}c<{\hspace{.2cm}}|c|>{\hspace{.2cm}}c<{\hspace{.2cm}}|c|c}
\hline
\multirow{2}{*}{Datasets} & \multicolumn{2}{c|}{$w_0-w_a$} & \multicolumn{4}{c}{FDS} \\
\cline{2-7}
 &  $\Delta\chi_{\rm MAP}^2$ & $\sigma$ & $\Delta\chi_{\rm MAP}^2$ & $\sigma$ & $c$ & $c'$ \\
\hline
\multirow{2}{*}{CMB} & \multirow{2}{*}{$-2.7$} & \multirow{2}{*}{1.1} & \multirow{2}{*}{0.0} & \multirow{2}{*}{0.0} & $-0.22$ & $0.01$ \\
 & & &  & & $-0.88^{+0.88}_{-0.30}$ & $0.041^{+0.012}_{-0.041}$ \\ 
\arrayrulecolor{gray!30}\hline
\multirow{2}{*}{CMB + DESI DR2 BAO} & \multirow{2}{*}{$-12.0$} & \multirow{2}{*}{3.0} & \multirow{2}{*}{-8.8} & \multirow{2}{*}{$2.5$} & $-1.03$ & $0.05$ \\
 &  & &  & & $-0.72^{+0.45}_{-0.43}$ & $0.048^{+0.012}_{-0.007}$ \\ 
\arrayrulecolor{gray!30}\hline
\multirow{2}{*}{CMB + DESI DR1 FS} &  \multirow{2}{*}{$-9.5$} & \multirow{2}{*}{$2.6$} & \multirow{2}{*}{$-10.3$} &\multirow{2}{*}{2.8} & $-0.74$ & $0.05$ \\
 &  & & & & $-0.58^{+0.36}_{-0.31}$ & $0.059^{+0.011}_{-0.009}$\\ 
 \arrayrulecolor{gray!30}\hline
\multirow{2}{*}{CMB + DESI1.5} & \multirow{2}{*}{$-11.1$} & \multirow{2}{*}{2.9} &  \multirow{2}{*}{$-14.6$} & \multirow{2}{*}{$3.4$} & $-0.82$ & $0.05$ \\
 &   & &  &  & $-0.61^{+0.20}_{-0.32}$ & $0.051^{+0.008}_{-0.006}$  \\ 
 \arrayrulecolor{gray!30}\hline
\multirow{2}{*}{CMB + DESI1.5 + DES-Dovekie} &  \multirow{2}{*}{$-16.1$} & \multirow{2}{*}{$3.6$} & \multirow{2}{*}{$-20.1$} & \multirow{2}{*}{$4.1$} & $-0.85$ & $0.05$ \\
 &   &  &  & & $-0.82_{-0.15}^{+0.11}$ & $0.053^{+0.007}_{-0.006}$  \\ 
\arrayrulecolor{black}\hline
\end{tabular}
\caption{Statistical significance comparing the CPL parametrization and the fading dark sector model across various combinations of datasets, to $\Lambda$CDM. Additionally, the best fit, the mean, and the $\pm 1\sigma$ values for $c,c'$ are provided for the fading dark sector model.}
\label{sigmas}
\end{table*}

\begin{table*}[!t]
\centering
\begin{tabular}{c|c|>{\hspace{.2cm}}c<{\hspace{.2cm}}|c|>{\hspace{.2cm}}c<{\hspace{.2cm}}|c|c}
\hline
\multirow{2}{*}{Datasets} & \multicolumn{2}{c|}{$w_0-w_a$} & \multicolumn{4}{c}{FDS} \\
\cline{2-7}
 &  $\Delta\chi_{\rm MAP}^2$ & $\sigma$ & $\Delta\chi_{\rm MAP}^2$ & $\sigma$ & $c$ & $c'$ \\
\hline
\multirow{2}{*}{CMB + DESI DR2 BAO + DES-Dovekie} & \multirow{2}{*}{$-13.7$} & \multirow{2}{*}{3.1} &  \multirow{2}{*}{$-13.2$} & \multirow{2}{*}{$3.2$} & $-0.92$ & $0.05$ \\
 &   & &  &  & $-0.82_{-0.20}^{+0.13}$ & $0.048^{+0.011}_{-0.007}$  \\ 
 \arrayrulecolor{gray!30}\hline
\multirow{2}{*}{CMB + DESI1.5 + DES-Dovekie} &  \multirow{2}{*}{$-16.1$} & \multirow{2}{*}{$3.6$} & \multirow{2}{*}{$-20.1$} & \multirow{2}{*}{$4.1$} & $-0.85$ & $0.05$ \\
 &   &  &  & & $-0.82^{+0.11}_{-0.15}$ & $0.053^{+0.007}_{-0.006}$  \\ 
\arrayrulecolor{black}\hline
\end{tabular}
\caption{The effect of adding DESI DR1 FS data to the dataset combination involving CMB + DESI DR2 BAO + DES-Dovekie on the statistical significance of the best-fit for the CPL parameterization and the fading dark sector model, compared to $\Lambda$CDM.}
\label{sigmas SN}
\end{table*}

\begin{table*}[!t]
\centering
\renewcommand{\arraystretch}{1.5}
\begin{tabular}{c|>{\hspace{.2cm}}c<{\hspace{.2cm}}|>{\hspace{.2cm}}c<{\hspace{.2cm}}|>{\hspace{.2cm}}c<{\hspace{.2cm}}|>{\hspace{.2cm}}c}
\hline
Datasets & $\Omega_m$ & $H_0$ & $\sigma_8$ & $n_s$\\
\hline
CMB & $0.326^{+0.024}_{-0.036}$ & $66.59^{+3.16}_{-3.10}$ & $0.810^{+0.029}_{-0.033}$ & $0.964\pm 0.004$\\
\arrayrulecolor{gray!30}\hline
CMB + DESI DR2 BAO & $0.312^{+0.007}_{-0.014}$ & $67.72^{+1.38}_{-0.80}$ & $0.820^{+0.014}_{-0.012}$ & $0.964\pm 0.004$\\
\arrayrulecolor{gray!30}\hline
CMB + DESI DR1 FS & $0.300\pm 0.010$ & $69.36^{+0.80}_{-1.41}$ & $0.842^{+0.012}_{-0.015}$ & $0.965\pm {0.004}$\\
 \arrayrulecolor{gray!30}\hline
CMB + DESI1.5 & $0.306_{-0.005}^{+0.006}$ & $68.47^{+0.41}_{-0.69}$ & $0.833\pm 0.010$ & $0.964\pm 0.004$\\
\arrayrulecolor{gray!30}\hline
CMB + DESI1.5 + DES-Dovekie & $0.310\pm {0.003}$ & $68.04^{+0.24}_{-0.38}$ & $0.829\pm {0.008}$ & $0.964^{+0.004}_{-0.003}$\\
\arrayrulecolor{black}\hline
\end{tabular}
\caption{Constraints on cosmological parameter of the FDS model. }
\label{cosmological parameters}
\end{table*}

Since the analysis pipeline we have developed involves new kernels and modifications of standard codes, we start by checking the outputs of these codes against known results. In particular, we validate the $\Lambda$CDM limit (i.e. $c'\rightarrow 0$ and $c\rightarrow 0$) against results published by the DESI collaboration. As mentioned in the appendix, our new kernels also reduce to the standard $\Lambda$CDM ones~\cite{Bernardeau:2001qr} in the appropriate limit. Our implementation is therefore under control in this limit.\footnote{The standard \texttt{CLASS-PT} implementation relies on the Einstein--de Sitter kernels which we have modified to be the exact-time kernels. We have additionally verified and reproduced the known percentage difference between the exact-time kernels compared to the Einstein--de Sitter kernels found in~\cite{Donath:2020abv,DAmico:2020tty}.}

Having checked the $\Lambda$CDM limit, we now allow the parameters $(c,c')$ to vary in an MCMC simulation to determine the preference for departures from $\Lambda$CDM. These parameters are added to the standard six $\Lambda$CDM parameters so our model has two extra parameters and we take that into account when calculating improved significances. In addition to MCMCs, we run minimizers to find the best fit parameter values. The result of these MCMC and minimizer runs are shown in Table~\ref{sigmas} where we compare our FDS model to the benchmark $w_0$--$w_a$ parametermization. The constraints on cosmological parameters $(\Omega_m,H_0,\sigma_8,n_s)$ in the FDS model from using various dataset combinations are shown in Table~\ref{cosmological parameters}. In order to better understand where constraints come from, we study different combinations of datasets to diagnose what information is added by each dataset. 

First, we study a baseline data variation where we include only CMB data as discussed in section~\ref{sec:additional data}. The CMB alone shows no preference for departures from $\Lambda$CDM. In particular, our model modifies late-time cosmology and leaves early-time CMB physics unaltered. For example, the scalar field has not changed appreciably from its initial value before recombination. This leaves the characteristic sound horizon scale $r_d$ fixed so that the CMB is mainly sensitive to integrated quantities along the line of sight. While there is additional information in the CMB dataset about late-time cosmology from late ISW and lensing, it seems like these effects are comparatively weak and do not show a preference for departures from $\Lambda$CDM. The mild improvement of fit, $\Delta\chi^2 \approx -2.7$ for $w_0$--$w_a$ is consistent with fitting noise with two extra parameters. In particular, the CPL parameterization has the freedom to wander into unphysical parameter space where secondary effects (such as the CMB lensing anomaly) fit slightly better.

Once the DESI BAO data is included, we break the degeneracy and get sensitivity to more than just integrated quantities to the CMB. In particular, DESI BAO provides information about distances to various intermediate redshifts between today and the CMB allowing different expansion histories that give the same distance to the CMB to be distinguished. With the inclusion of BAO data, we see a preference for $w_0$--$w_a$ and FDS over $\Lambda$CDM that is ultimately rooted in the $\Omega_m$ tension described in~\cite{DESI:2025zgx}. Since the included BAO data is based on background distances, this preference is driven by departures of the distances measured by DESI from those predicted by the CMB within $\Lambda$CDM. This raises the question of whether measurements of the growth of structure (without the BAO peak) also agree with this conclusion.

We answer this question by studying the full-shape (FS) power spectrum of DESI DR1 in combination with the CMB. The FS data contains weaker information about the BAO scale which DESI provides by a reconstruction procedure that undoes the effects of bulk flows. On the other hand, FS data contains other information that can be obtained from redshift-space distortions (RSD), the broadband shape of the spectrum and the Alcock--Paczynski effect. As mentioned previously, these are sensitive to the growth of structure and also provide information on cosmological parameters affecting the expansion history of the universe. As seen in Table~\ref{sigmas}, we find that the FS data has the same preference as DESI DR2 BAO with deviations at less than $1\sigma$. The FDS model provides a marginally better fit to CMB+FS compared to $w_0$--$w_a$ although the difference in $\chi^2$ is only about 0.8 which could be due to noise. This is because FDS can match the low $\Omega_m$ preferred by DESI data without reducing the amplitude of structure while in $w_0$--$w_a$ there is no independent boost to the amplitude of structure. Moreover, FS data provides a better measurement of the parameter $c$, reducing the error by a factor of 1.3. From the constraints in figure~\ref{fig:ccp} and in table~\ref{sigmas}, we see that FS data prefers a larger value of $c'$ compared to DESI BAO. However, late-time effects depend on the product $cc'$ since the field $\phi$ rolls due to its potential with logarithmic slope $c$ and the DM fading is controlled by $c'$. In order to prevent effects associated to the fading of DM and modifications to the background cosmology from becoming too big, the fit prefers smaller values of $|c|$, thereby also tightening the error bars. 

Since the modified $w_0$--$w_a$ and FDS are consistent under the BAO and FS datasets separately, we study the combination of both datasets which we call DESI1.5. With this combination, we see a remarkable reversal of preference for FDS over $w_0w_a$ by about $\Delta\chi^2 \approx -3.5$. The three data combinations suggest that $w_0w_a$ provides a slightly better fit to background distances measured by DESI but the preference tips in favor of FDS once the FS power spectrum is included. This indicates that BAO and FS data favor different regions of the $w_0$--$w_a$ parameter space, whereas for the FDS model the two datasets reinforce each other. With DESI1.5, we again see a tightening of the errors on $c$ by a factor of 1.7 compared to CMB+DESI DR2 BAO. 

Finally, we include the supernovae (SNe) sample from DES-Dovekie which is a measure of the expansion history at low redshift. When combined with CMB+DESI DR2 BAO, the SNe sample doesn't distinguish between FDS and $w_0$--$w_a$~\cite{Bedroya:2025fwh}. However, with FS data the FDS model reaches a $4.1\sigma$ preference over $\Lambda$CDM and is favored over $w_0$--$w_a$ by $\Delta\chi^2 \approx -4$. This is further highlighted in Table~\ref{sigmas SN}. This is a similar preference to the one mentioned above without SNe. We see that the SNe do not disfavor the region of parameter space selected by CMB+DESI1.5 and in this sense the datasets all reinforce each other. The inferred values of $(c,c')$ also remain consistent with the ones from CMB+DESI1.5. The parameter $c$ shifts to slightly more negative values reflecting the preference of the SNe data for larger $\Omega_m$. This is an illustration of the role played by the potential in the FDS model. As observed in~\cite{Clini:2026kdj}, a model with a dark fifth force would have difficulty accommodating the SNe data but we see here that our model is capable of doing that precisely because the mediator of the fifth force also plays the role of dark energy. Future surveys such as Euclid~\cite{Euclid:2024yrr} or SNe samples from ZTF~\cite{Rigault:2024kzb}, Rubin~\cite{LSSTDarkEnergyScience:2018jkl} and Roman~\cite{Kessler:2025eib} will provide more constraining power and may be able to distinguish between models such as FDS and $w_0$--$w_a$ more decisively. 

\subsection{Growth of structure}

\begin{figure*}[!tp]
    \centering
    \begin{subfigure}{.49\textwidth}
    \includegraphics[width=\linewidth]{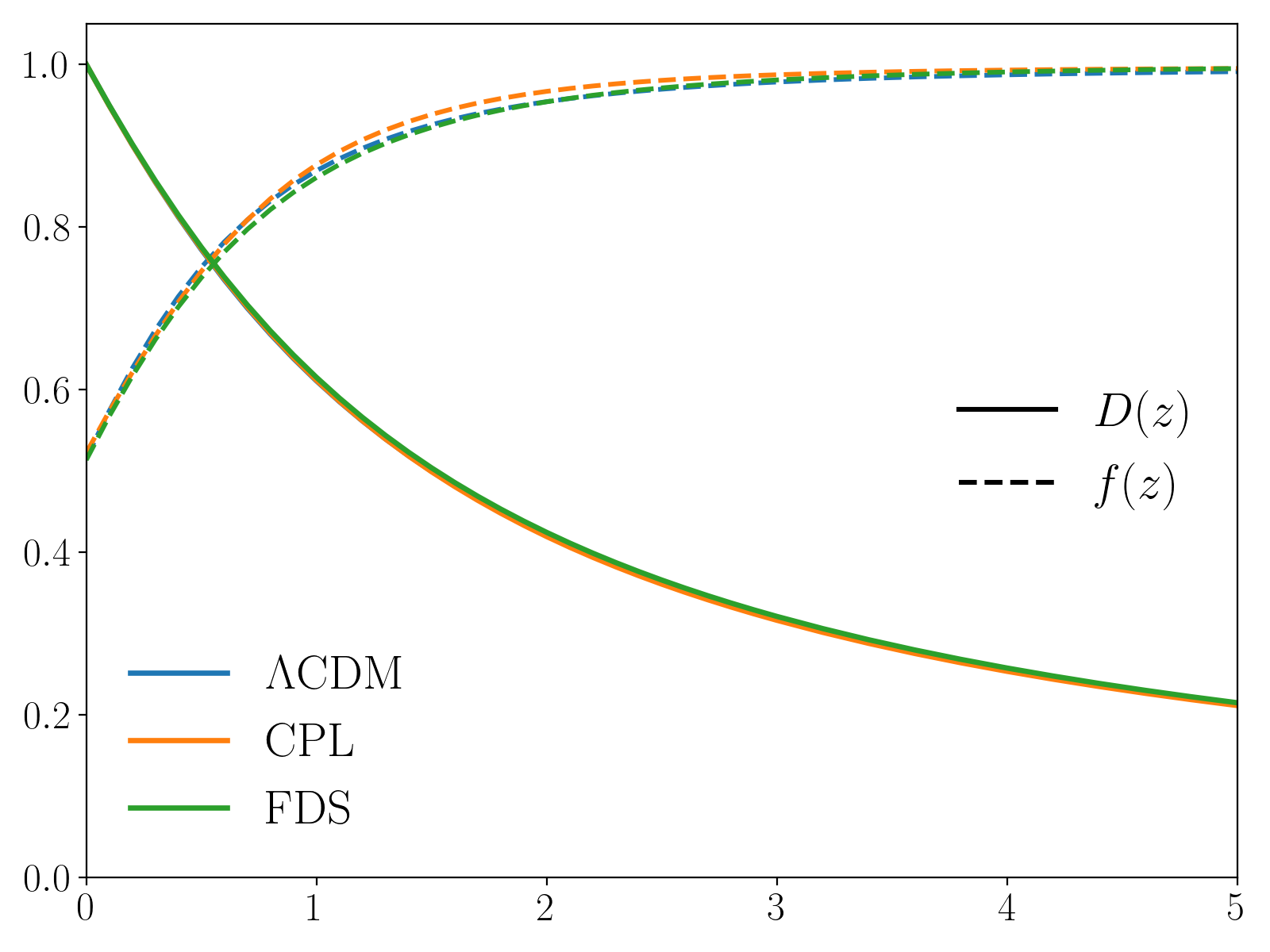}
    \begin{picture}(0,0)
        \put(0,5){$z$}
    \end{picture}\vspace*{-0.5cm}
    \caption{Growth functions}
    \end{subfigure}
    \hfill
    \begin{subfigure}{.49\textwidth}
    \includegraphics[width=\linewidth]{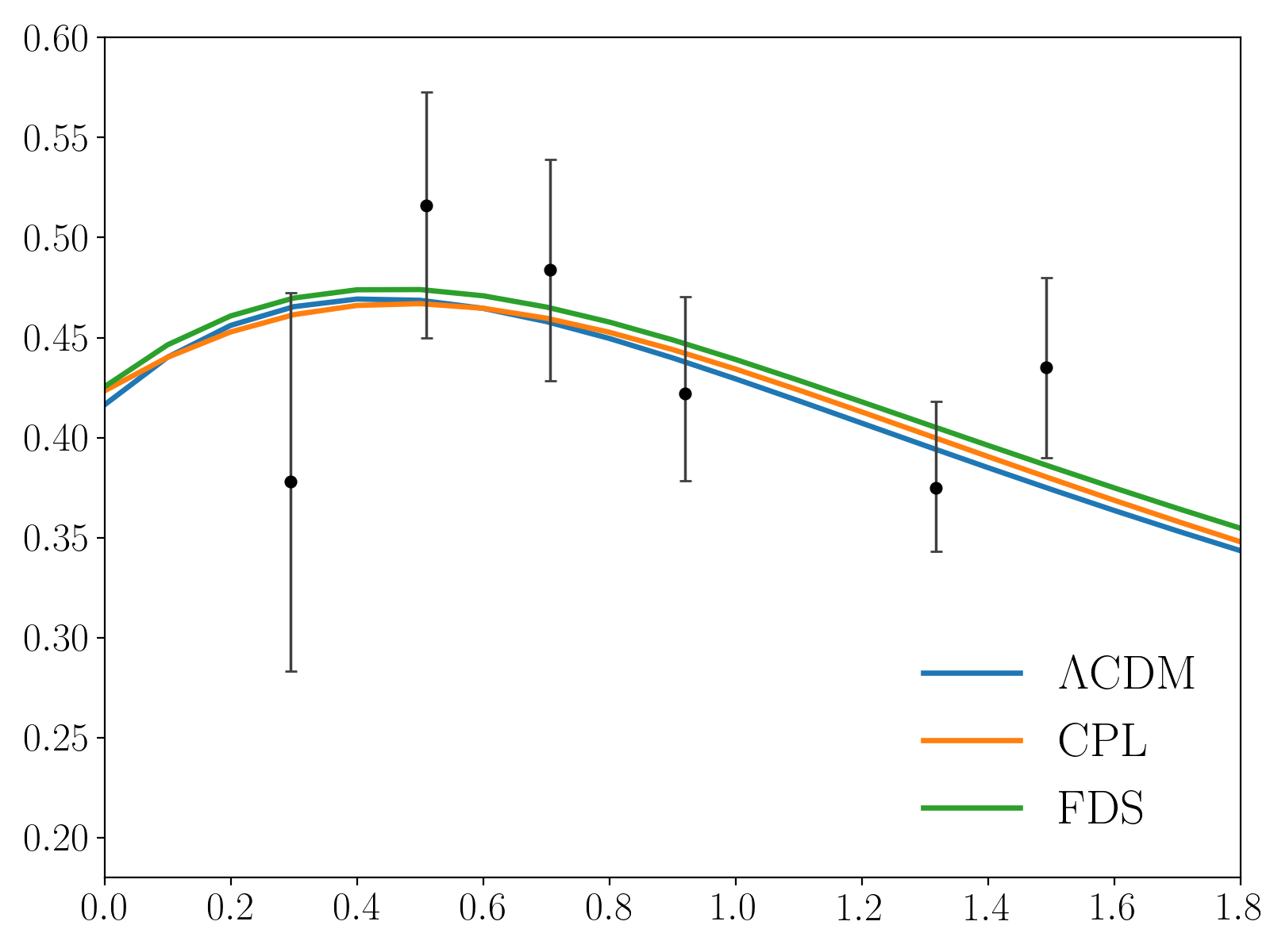}
    \begin{picture}(0,0)
        \put(0,5){$z$}
        \put(-130,85){\rotatebox{90}{$f(z)\sigma_8(z)$}}
    \end{picture}\vspace*{-0.5cm}
    \caption{Growth of structure}
    \end{subfigure}
    \caption{Comparison among the best-fit growth factor $D(z)$ and growth rate $f(z)$ of $\Lambda$CDM, the CPL parameterization, and the FDS model at $k=0.03\, h\,\mathrm{Mpc}^{-1}$ from fitting to CMB+DESI1.5+DES-Dovekie. The growth factor $D(z)$ is normalized to its value today.}
    \label{fig:growth}
\end{figure*}

The FS data favor the FDS model because of its altered growth of structure compared to $w_0$--$w_a$. In this section, we examine this preference in more detail. We start by recalling the definition of the linear growth factor $D(z)$ and the linear growth rate $f(z)$:
\begin{align}
    D(z)&=\sqrt{P_L(k,z)/P_L(k,0)} \\
    f(z)&=\frac{\dd\ln D(k,z)}{\dd \ln a}.
\end{align}
These curves are shown in figure~\ref{fig:growth} for the models we consider in this analysis. In principle, $D$ and $f$ depend on the scale $k$ through the range of the fifth force but on scales relevant for the DESI data, we have $k \gg a m_{\rm eff}$ and the dependence is very weak. This is illustrated in figure~\ref{fig:strength}. Because of this we choose to evaluate $D$ and $f$ at $k = 0.03 h/\rm{Mpc}$ which is the pivot for the ShapeFit analysis~\cite{DESI:2024jxi}. The differences in the growth functions between models are due to slightly different expansion histories as well as evolution of perturbations. 

In the context of FDS, The growth of structure is different because of competing physical effects. On the one hand, there is a fifth force in the dark sector whose strength is proportional to $2c'^2$ and this adds to the gravitational attraction and speeds up matter in-fall. Additionally, as dark matter loses mass, momentum dilution due to the cosmological expansion becomes slower and further enhances the growth of structure. On the other hand, the fading of dark matter decreases its mass and slows down gravitational collapse thereby reducing the growth. The importance of these effects varies during cosmic evolution and in some eras they may even be reversed (e.g. the dark matter gets heavier during the dark energy era in our model). The net effect of these new phenomena is conveniently summarized by the effective growth index. 

The effective growth index $\gamma_{\rm eff}$ expresses the linear growth rate in terms of the fractional matter density~\cite{Linder:2005in}:
\begin{align}
    f(z)=\Omega_m(z)^{\gamma_{\rm eff}(z)}
\end{align}
where $\Omega_m(z)$ is the actual matter fraction including the effects of fading. In our model we find $\gamma_{\rm eff} \approx 0.57$ for $z\ll 1$ compared to the standard value $\gamma_{\Lambda{\rm CDM}} = 0.55$ in $\Lambda$CDM. Since $\Omega_m(z)$ is smaller than 1 at late times, the larger value of $\gamma_{\rm eff}$ means that structure grows more slowly in our model than $\Lambda$CDM. DESI's measured growth index gives $\gamma_{\rm DESI} \approx 0.58 \pm 0.11$~\cite{Turner:2025xpy} is consistent with both values although the central value although the central value slightly prefers a suppressed growth rate, as in our model. 

Observations, such as redshift-space distortions, provide a more direct measurement of the quantity $f\sigma_8(z)$, i.e. the product of the linear growth rate and amplitude of matter fluctuations at  8 $h^{-1}{\rm Mpc}$ computed from the linear power spectrum. We show a plot of $f\sigma_8$ from DESI's ShapeFit analysis~\cite{DESI:2024jxi} in figure~\ref{fig:growth} along with theoretical curves for the models studied in this paper. The data points are merely illustrative because they have been obtained with templates assuming a $\Lambda$CDM cosmology. They are not used directly in the likelihood. The theoretical curves show that FDS has slightly more structure than CPL and $\Lambda$CDM. This is to be contrasted with the slightly slower growth rate mentioned previously. These two facts imply that the extra physical effects (such as fifth forces) in FDS acting through the early history of the universe build a larger $\sigma_8$ amplitude even though the growth rate today is slightly slower. Observations seem to prefer this combination of larger amplitude and slower growth rate which our FDS model is able to accommodate.

\section{Discussion}
\label{sec:discussion}

In this paper, we studied the Fading Dark Sector (FDS) including data from the full shape (FS) of the galaxy power spectrum measured by DESI in their first data release. We found that adding the FS data (without including any SN dataset), raises the significance of our model from $2.5\sigma$ to $3.4\sigma$ while keeping the inferred value of $c'$ stable at around $0.05\pm 0.01$. The FS data also tightens our measurement of $c$, reducing the error bar by a factor of $1.3 - 1.7$. This improvement comes from the sensitivity of the FS analysis to the growth of structure which proceeds differently in our model due to the presence of fifth forces and varying dark matter masses.

The importance of these additional features of the FDS model is most visible when the model is compared to the standard cosmological alternative, i.e. the $w_0$--$w_a$ parameterization. While the DESI BAO data has a slight preference for $w_0$--$w_a$ at the level of $0.5\sigma$, the situation is reversed once the FS data is included. This is an indication that $w_0$--$w_a$ lacks the features necessary to provide a good fit to the growth of structure, even though it has more flexibility to fit the background cosmology by venturing into unphysical regions of parameter space. For the FDS model, the altered growth of structure is a prediction since it is determined by the same $(c,c')$ parameters that fix the background evolution. From this perspective, FDS passes a non-trivial check since $(c,c')$ remain unchanged with additional data. Moreover, the improvement in significance of the FDS model with additional data indicates that these features are important for cosmology. This is unlike the $w_0$--$w_a$ parameterization which is disfavored compared to FDS by the inclusion of additional FS data. 

Other models in the literature such as~\cite{Bottaro:2023wkd,Clini:2026kdj} also have dark sector fifth forces. Analyses of these models have provided hints for a dark fifth force consistent in strength with what we find. However, our FDS model includes additional dynamics in the dark sector because of the potential $V(\phi)$ and these dynamics are favored by data at the level of $\Delta\chi^2 = -20.1$. When including DES-Dovekie data (we expect the same for other SNe data \cite{Bedroya:2025fwh} but have not run those with FS analysis), the error bar on $c$ is tightened while remaining consistent with previous results. Without the dynamics  of dark energy dictated by $V(\phi)$, it is unclear whether a model would be able to accommodate the SNe data~\cite{Clini:2026kdj}. The value of $c'$ remains consistent under all cosmological probes predicting a dark fifth force $2c'^2 \approx 0.5\%$ the strength of gravity in the dark sector. Galactic dynamics can offer an independent test of this force.

Recently, studies of U-shaped galactic disk warps have placed stringent constraints on the strength of a dark fifth force bounding $\Delta G_N/G_N \lesssim 1.7 \times 10^{-3}$ \cite{Desmond:2026lds}. This result is in tension with coupled dark sector resolutions to the DESI-CMB-SNe tension and would rule out our value of $c'\approx 0.05$. 
However that paper assumes a core density profile which may be different for our model because of the dark matter decay and the resulting kick velocity which partially flattens the core as we show in an upcoming work~\cite{Fulcher26}.  In addition the baryonic activity at the core may contribute to further flattening of the density profiles at the core.  The quoted limit is conditional on the halo-profile prescription and in particular its cuspy core and it would be important to check the robustness of this limit to observationally allowed broad cores. This, as well as other warp-modeling systematics, could weaken this bound making it compatible with our model prediction of a long range dark fifth force around $0.5\%$ of the gravitational force. 
Resolving this tension would require further scrutiny of the observational and theoretical methods involved. 

Our analysis could be improved in various ways. First, we did not make use of the available galaxy bispectrum data but earlier results (in the context of $\Lambda$CDM and CPL, see e.g.,~\cite{Chudaykin:2025aux,Forero-Sanchez:2026zfb,Chudaykin:2025lww,Novell-Masot:2026sgg}), 
show only small deviations when the bispectrum data is included. We believe this will be the case for our model as well so we leave the inclusion of the bispectrum data for the future. Moreover, models of coupled dark sectors have not been validated to the same extent as $\Lambda$CDM in the context of the EFTofLSS. That is to say, one could run N-body simulations with a coupled dark sector to test the robustness of the inference process and check whether the cosmology could be reproduced faithfully. Since our model is a small deviation from $\Lambda$CDM, we believe that qualitative results along this direction in the context of $\Lambda$CDM should carry over to FDS. 

In the near future, several datasets will further test the FDS hypothesis. The most immediate is DESI DR2 FS power spectrum which will be a more precise measurement of the growth of structure. Beyond DESI, Euclid~\cite{Euclid:2024yrr} and Roman~\cite{Wang:2021oec} will map galaxy clustering and also measure the growth of structure, providing further tests of the FDS model. Supernova observations will also improve the constraints on the expansion history. These include samples from ZTF~\cite{Rigault:2024kzb} and the Vera Rubin Observatory~\cite{LSSTDarkEnergyScience:2018jkl} which will extend the number and redshift of observed supernovae. Since our potential parameter $c$ is constrained mainly by the late-time expansion, these data will play a decisive role in our understanding of dynamical dark energy.

In summary, cosmological distance measurements provided hints that something beyond $\Lambda$CDM is needed, but several models are degenerate under these measurements alone. The full shape of the galaxy power spectrum begins to distinguish between these models. FDS is a model that contains two additional parameters which can be determined from distance measurements and these same inferred values turn out to account for the growth of structure outperforming the $w_0$--$w_a$ parameterization by $\Delta \chi^2 \approx -4$ for the same number of parameters. When SNe data is included, the model reaches a $4.1\sigma$ preference over $\Lambda$CDM and is, to our knowledge, the only model in the literature with this high significance, in addition to being motivated from string theoretic considerations. If upcoming data confirm this departure from $\Lambda$CDM, the dynamics of the dark sector would imply the presence of a fifth force that cosmology and galactic surveys can measure.

\subsection*{Acknowledgements}

We would like to thank Mustapha Ishak-Boushaki for encouraging us to do the full-shape analysis and for helpful conversations.  We would also like to thank the SCGP for the 2026 summer workshop in physics where this work was initiated.

The simulations in this paper were performed on the FASRC Cannon cluster supported by the FAS Division of Science Research Computing Group at Harvard University. The implementation of the FDS model is available at \href{https://github.com/dhwu14/CLASS-PT-exp}{\faGithub\ \texttt{dhwu14/CLASS-PT-exp}}.
We used OpenAI ChatGPT-6 to assist with coding and interfacing \verb+CLASS-PT+ with DESI DR1 FS data, which have all been checked and verified by the authors. We take full responsibility for the results presented in this paper.

The work of CV and DW is supported in part by a grant from the Simons Foundation (602883, CV) and the DellaPietra Foundation. 

\bibliography{apssamp}

\onecolumngrid

\appendix

\section{Effective field theory for the fading dark sector}
\label{app:eft_derivation}

In this appendix, we derive the perturbative equations associated to the fading dark sector. The background and linear perturbation equations of motion have been derived for coupled dark sectors in~\cite{Amendola:1999dr,Miranda:2017rdk}. The fading dark sector has been first implemented in \verb+CLASS+ in~\cite{Agrawal:2019dlm} and subsequently studied using recent cosmological data~\cite{Bedroya:2025fwh}. Here, we focus on the higher-order corrections coming from the one-loop corrections computed using the EFTofLSS for the fading dark sector model. The computations are implemented as a modification of \verb+CLASS-PT+~\cite{Chudaykin:2020aoj}. We denote the fading dark matter species by $\chi$ in order to avoid confusion with the potential parameter $c$ (appearing as $V(\phi)=V_0e^{-c\phi}$). We denote baryons by $b$. Additionally, we denote $H=\dot{a}/a$, $\cH=aH$, primes as derivatives with respect to conformal time, and introduce $N=\ln a$ for convenience. We also work in Planck units with $M_\pl=(8\pi G_N)^{-1/2}=1$. We also adopt the Fourier normalization as $F(x)=\frac{1}{(2\pi)^3}\int \dd^3ke^{i\mathrm{k}\cdot\mathrm{x}}F(\mathrm{k})$. The perturbative expansion is done in the pressureless regime with metric perturbations treated at the sub-horizon Newtonian order. The scalar response in the nonlinear kernels is quasi-static and linear in the scalar fluctuation. 

We start with the scalar field background equation, which for the fading dark sector takes on the following form
\begin{equation}
    \phi''+2\cH\phi'-a^2cV(\phi)-a^{-1}c'\rho_{\chi,0}e^{-c'\phi}=0\,.
\end{equation}
The equation of motion for the linear perturbation of the scalar field is then
\begin{equation}
    \delta\phi''+2\mathcal{H}\delta\phi'+\left(k^2+a^2m_{\rm eff}^2\right)\delta\phi=a^2c'\rho_\chi\delta_\chi-\frac12 \phi'h'\,,
\end{equation}
where $\delta_\chi=\delta n_\chi/n_\chi$ with $\rho_\chi=n_\chi m_{\rm DM}(\phi)$, $h$ denotes the metric perturbation in synchronous gauge and the effective scalar mass is\footnote{Recall in synchronous gauge, to linear order, the metric is 
$$
g^{00}=-\frac{1}{a^2}\,,\qquad g^{ij}=\frac{1}{a^2}\big(\delta^{ij}-h^{ij}\big)\,,\qquad \sqrt{-g}=a^4\bigg(1+\frac{h}{2}\bigg)\,.
$$
}
\begin{equation}
\label{eq:meff}
    m_{\rm eff}^2=c^2V+c'^2\rho_\chi\,.
\end{equation}
The dark matter density behaves as $\rho_{\chi}=\rho_{\chi,0}a^{-3}e^{-c'\phi}$.

For $k\gg\mathcal{H}$ and $k^2+a^2m_{\rm eff}^2\gg\mathcal{H}^2$, we work in the sub-horizon, pressureless, quasi-static (QS) regime where the scalar perturbation reduces to
\begin{equation}
\label{eq:dphi_qs}
    \delta\phi(\bk,a)\simeq\frac{a^2c'\rho_\chi}{k^2+a^2m_{\rm eff}^2}\delta_\chi(\bk,a)\,.
\end{equation}
The scalar field therefore produces an attractive force acting on dark matter, with the physical fifth-force enhancement being $\Gamma$, as denoted in the main text in \cref{eq:main_Gamma}.

Let us next derive the perturbation equations for the fading dark matter and baryon fluids in Newtonian gauge. The number current for each conserved species obeys $\nabla_\mu (n_Iu_I^\mu)=0$ where $I\in\{b,\chi\}$ represents the species index. In the non-relativistic limit, with $n_I(\mathbf{x},\tau)=n_I(\tau)[1+\delta_I(\mathbf{x},\tau)]$ and $n_I'+3\cH n_I=0$, we arrive at the continuity equation~\cite{Ma:1995ey}
\begin{equation}
    \delta_I'+\theta_I=-\nabla\cdot (\delta_I\mathbf{v}_I)\,,
\end{equation}
where $\mathbf{v}_I=\dd \mathbf{x}/\dd \tau$. The standard linear relativistic equations for baryons and dark matter moving under the Newtonian potential $\Psi$ can be found in~\cite{Ma:1995ey}. The Newtonian gravitational potential obeys $\nabla^2\Psi=a^2\delta\rho/2$ and, to first order in $\Psi$, the spacetime metric is
\begin{equation}
    \dd s^2=-(1+2\Psi)\dd t^2+a^2(t)(1-2\Psi)\dd \mathbf{x}^2\,.    
\end{equation}
For the coupled dark matter, differentiating $\ln m_\chi$ gives an additional velocity drag term and the scalar acceleration. 
Hence, the Euler equations in the $k\gg \cH$ limit and deep in the matter domination era can be approximated as 
\begin{align}
    \mathbf{v}_b'+\cH\mathbf{v}_b+(\mathbf{v}_b\cdot \nabla)\mathbf{v}_b&=-\nabla \Psi\,,\\
    \mathbf{v}_\chi'+(\cH-c'\phi')\mathbf{v}_\chi+(\mathbf{v}_\chi\cdot \nabla)\mathbf{v}_\chi&=-\nabla\Psi+c'\nabla\delta\phi\,,
\end{align}
where the sub-horizon gravitational potential obeys
\begin{equation}
    -k^2\Psi=\frac{a^2}{2}\left(\rho_b\delta_b+\rho_\chi\delta_\chi\right)\,.
\end{equation}
Further taking the divergence of the Euler equations, we obtain
\begin{align}
\label{eq:baryon Euler}
    \theta_b'+\cH\theta_b&=-\frac{a^2}{2}\delta\rho-\nabla\cdot[(\mathbf{v}_b\cdot\nabla)\mathbf{v}_b]\,,\\
    \theta_\chi'+(\cH-c'\phi')\theta_\chi&=-\frac{a^2}{2}\delta\rho+c'\nabla^2\delta\phi-\nabla\cdot[(\mathbf{v}_\chi\cdot\nabla)\mathbf{v}_\chi]\,,
\end{align}
where $\theta_I=\nabla\cdot\mathbf{v}_I$. 
Notice that \eqref{eq:baryon Euler} contains no pressure $c_s^2k^2\delta_b$ and no tight coupling between photons and baryons $(4\rho_\gamma/3\rho_b)an_e\sigma_T(\theta_\gamma-\theta_b)$ appearing in the usual Newtonian gauge~\cite{Ma:1995ey}. For the pressure term, we can compare with the gravitational potential which upon a Fourier transformation is
\begin{equation}
    \frac{|c_s^2k^2\delta_b|}{|k^2\Psi|}=\frac{2}{3\Omega_m(a)}\bigg(\frac{c_s k}{\cH}\bigg)^2\bigg|\frac{\delta_b}{\delta_m}\bigg|\,.
\end{equation}
The (baryonic) Jeans scale can be deduced when we balance the pressure and the gravitational term leading to $c_s^2k_J^2=\frac{3\Omega_m(a)}{2}\cH^2$.\footnote{For a non-relativistic ideal baryon gas, the sound speed is $c_s^2\sim k_BT_b/(\mu m_p c^2)\ll 1$ where $\mu m_p$ denotes the mean particle mass. However $(ck/\cH)^2\gg 1$ on subhorizon scales. Thus, non-relativistic baryons do not guarantee the pressure is negligible.} Hence, in the linear and quasi-linear regime of interest to the EFTofLSS with $k\ll k_J$, the pressure term becomes negligible. On the other hand, within the photon baryon coupling term, the $an_e\sigma_T$ denotes the Thomson scattering rate per unit conformal time while  $4\rho_\gamma/3\rho_b$ denotes the conversion of photon momentum exchange into acceleration of baryons. In the tight-coupling era before recombination $(z\approx 1100)$, the term is important as the coupling is large $an_e\sigma_T\gg H$, even though $\theta_\gamma\sim \theta_b$. However, shortly after recombination, the free electron abundance decreases sharply, leading to the Thomson scattering rate to decrease and photons no longer exert a strong drag on the baryon fluid. We can compare this effect to the Hubble friction term 
\begin{equation}
    \frac{4\rho_\gamma}{3\rho_b}\frac{n_e\sigma_T}{H}\sim \frac{x_e(a)a^{-4}}{H(a)}\,,
\end{equation}
where $x_e(a)\sim n_ea^3$ denotes the ionized fraction. Deep in the matter domination era, $H(a)\sim a^{-3/2}$ such that the above ratio becomes $\sim x_e(a)a^{-5/2}$ which implies that the coupling term becomes negligible compared to its amplitude near recombination.

\begin{figure}[!tp]
    \centering
    \includegraphics[width=0.5\linewidth]{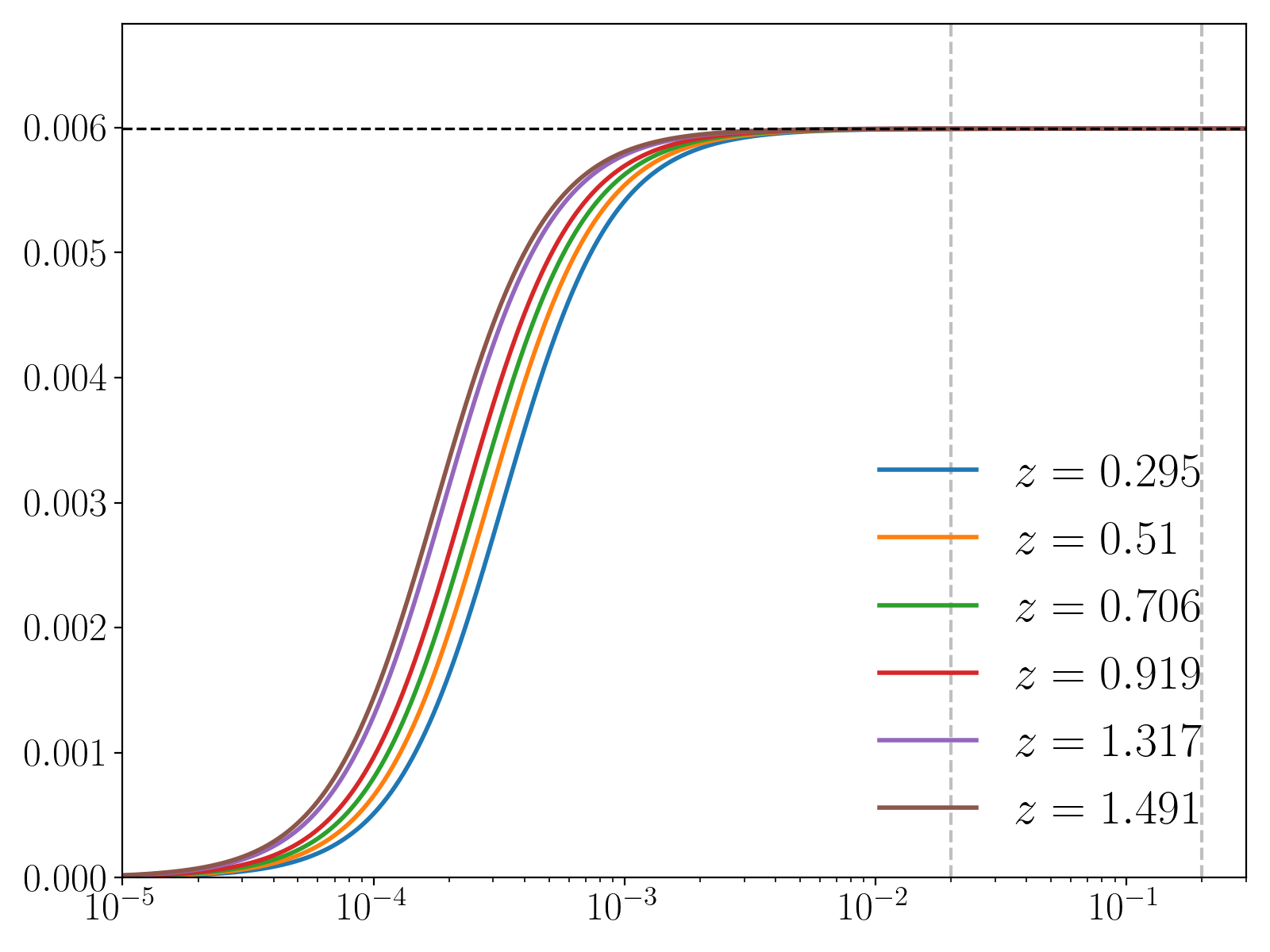}
    \begin{picture}(0,0)
        \put(-130,-5){$k\,[h\,\mathrm{Mpc}^{-1}]$}
        \put(-260,80){\rotatebox{90}{$\mu_\chi-1$}}
        \put(0,155){$2c'{}^2$}
    \end{picture}\vspace*{0.2cm}
    \caption{The behavior of the strength of the fifth force in the dark sector as a function of redshift $z$ and wavenumber $k$ for the best-fit FDS model from fitting to CMB + DESI1.5 + DES-Dovekie. The sampled redshifts are relevant to DESI DR1 while the gray dashed vertical lines indicate the range of $k$ to which DESI DR1 full-shape data measures. The black horizontal dashed line indicates the value of $2c'{}^2$ for this best-fit FDS model.}
    \label{fig:strength}
\end{figure}

When the large-scale flow is irrotational $\nabla\times \mathbf{v}_I=0$, we consequently have $(\mathbf{v}_I\cdot\nabla)\mathbf{v}_I=\nabla(\mathbf{v}_I^2/2)$. Setting $\bq$ as the velocity momentum, $\bp$ as the density momentum, and $\mathbf{k}=\mathbf{p}+\mathbf{q}$, the Fourier transform of the perturbation equations become
\begin{align}
\label{eq:pert fourier 1}
    \delta_I'(\bk)+\theta_I(\bk)&=-\int\frac{\dd^3q}{(2\pi)^3}\alpha(\bq,\bp)\theta_I(\bq)\delta_I(\bp)\,,\\
\label{eq:pert fourier 2}
    \theta_b'(\bk)+\cH\theta_b(\bk)&=-\frac{a^2}{2}(\rho_b\delta_b+\rho_\chi\delta_\chi)(\bk)-\int\frac{\dd^3q}{(2\pi)^3}\beta(\bq,\bp)\theta_b(\bq)\theta_b(\bp)\,,\\
\label{eq:pert fourier 3}
    \theta_\chi'(\bk)+(\cH-c'\phi')\theta_\chi(\bk)&=-\frac{a^2}{2}(\rho_b\delta_b+\rho_\chi\mu_\chi\delta_\chi)(\bk)-\int\frac{\dd^3q}{(2\pi)^3}\beta(\bq,\bp)\theta_\chi(\bq)\theta_\chi(\bp)\,,
\end{align}
where we have introduced 
\begin{equation}
    \mu_\chi(\bk)=1+2c'{}^2\frac{k^2}{k^2+a^2m_{\rm eff}^2}\,,\qquad \alpha(\bq,\bp)=\frac{(\bq+\bp)\cdot \bq}{q^2}\,,\qquad \beta(\bq,\bp)=\frac{|\bq+\bp|^2\bq\cdot\bp}{2q^2p^2}\,.
\end{equation}
The time dependence and the $k$-dependence of $\mu_\chi$ are shown in figure~\ref{fig:strength} where in the regime relevant to the DESI DR1 FS measurements, $\mu_\chi\approx 1+2c'{}^2$ remains a good approximation.

We can further transform these perturbative equations by changing $\dd/\dd \tau$ to $\dd/\dd N$ and $\theta=-\cH \Theta$. The continuity equation becomes
\begin{equation}
    \delta_{I,N}-\Theta_I=\int\frac{\dd^3q}{(2\pi)^3}\alpha(\bq,\bp)\Theta_I(\bq)\delta_I(\bp)\,.    
\end{equation}
To convert the Euler equations, we have
\begin{equation}
    \theta'=-\cH^2\bigg(\frac{\dd\Theta}{\dd N}+\frac{\dd\ln\cH}{\dd N}\Theta\bigg)\,.
\end{equation}
These lead to
\begin{align}
    \Theta_{b,N}&=-\bigg(2+\frac{\dd \ln H}{\dd N}\bigg)\Theta_b+\frac32\big(\Omega_b\delta_b+\Omega_\chi\delta_\chi\big)+\int\frac{\dd^3q}{(2\pi)^3}\beta(\bq,\bp)\Theta_b(\bq)\Theta_b(\bp)\,,\\
    \Theta_{\chi,N}&=-\bigg(2+\frac{\dd \ln H}{\dd N}-c'\phi_{,N}\bigg)\Theta_\chi+\frac32\big(\Omega_b\delta_b+\Omega_\chi\mu_\chi\delta_\chi\big)+\int\frac{\dd^3q}{(2\pi)^3}\beta(\bq,\bp)\Theta_\chi(\bq)\Theta_\chi(\bp)\,,
\end{align}
where we have made use of $\frac{a^2\rho_I}{2\cH^2}=\frac32\Omega_I$.

\subsection{Perturbation kernels and matter loops}

To solve the above equations of motion, we can expand both baryon and dark matter by a common Gaussian initial field $\delta_*(\bk)$ (see e.g.,~\cite{Bernardeau:2001qr} and references therein)
\begin{align}
\label{eq:kernel Fs}
    \delta_I^{(n)}(\bk,N)&=\int\bigg[\prod_{a=1}^n\frac{\dd^3 k_a}{(2\pi)^3}\delta_*(\bk_a)\bigg](2\pi)^3\delta_D^{(3)}\bigg(\bk-\sum_{a=1}^n\bk_a\bigg)F_{n,I}(\bk_1,\dots,\bk_n;N)\,,\\
\label{eq:kernel Gs}
    \Theta_I^{(n)}(\bk,N)&=\int\bigg[\prod_{a=1}^n\frac{\dd^3 k_a}{(2\pi)^3}\delta_*(\bk_a)\bigg](2\pi)^3\delta_D^{(3)}\bigg(\bk-\sum_{a=1}^n\bk_a\bigg)G_{n,I}(\bk_1,\dots,\bk_n;N)\,,
\end{align}
where $\delta_D$ denotes the Dirac delta function.
At first order, $F_{1,I}(\cK,N)$ and $G_{1,I}(\bk,N)$ are the density and normalized velocity-divergence growth functions and satisfy the following relations
\begin{equation}
    \frac{\dd}{\dd N}\begin{pmatrix}
        F_{1,b}\\
        G_{1,b}\\
        F_{1,\chi}\\
        G_{1,\chi}
    \end{pmatrix}
    =
    \mathbf{M}(k,N)
    \begin{pmatrix}
        F_{1,b}\\
        G_{1,b}\\
        F_{1,\chi}\\
        G_{1,\chi}
    \end{pmatrix}\,,
\end{equation}
where 
\begin{equation}
    \mathbf{M}(k,N)=\begin{pmatrix}
        0&1&0&0\\
        \frac32\Omega_b & -2-\frac{\dd \ln H}{\dd N} & \frac32 \Omega_\chi & 0\\
        0&0&0&1\\
        \frac32\Omega_b&0&\frac32\Omega_\chi\mu_\chi&-2-\frac{\dd \ln H}{\dd N}+c'\phi_{,N}
    \end{pmatrix}
\end{equation}
This is the exact linear operator on the non-linear quasi-static kernels. Unlike the Einstein--de Sitter kernels, the quasi-static kernels will generally depend on $k$ due to $\mu_\chi$. At linear order, a mode at $\bk$ evolves independently
\begin{equation}
    \delta_I^{(1)}(\bk,N)=F_{1,I}(k,N)\delta_*(\bk)\,,\qquad \Theta_I^{(1)}=G_{1,I}(k,N)\delta_*(\bk)\,.
\end{equation}
Thus, $F_{1,I}$ controls how the density perturbation evolves while $G_{1,I}$ determines how the velocity divergence evolves. In fact, these are correlated
\begin{equation}
    G_{1,I}(k,N)=\frac{\partial F_{1,I}(k,N)}{\partial N}\,.
\end{equation}
Furthermore, we can specify the initial condition on these kernels as
\begin{equation}
\label{eq:recursion}
    F_{1,I}(k,N_{\rm ini})=\frac{\delta_I^{\rm lin}(k,N_{\rm ini})}{\delta_m^{\rm lin}(k,N_{\rm ini})}\,,\qquad G_{1,I}=-\frac{\theta_I^{\rm lin}(k,N_{\rm ini})}{\cH_{\rm ini}\delta_m^{\rm lin}(k,N_{\rm ini})}\,.
\end{equation}
At higher orders, this same linear operator acts on each kernel, while lower-order products become inhomogeneous sources. Defining $\mathbf{K}:=(F_{n,b},G_{n,b},F_{n,\chi},G_{n,\chi})$, we obtain the time-dependent recursion relation for the fading dark sector model
\begin{equation}
    \frac{\dd}{\dd N}\mathbf{K}_{n}=\mathbf{M}\bigg(|\sum_{a=1}^n \bk_a|,N\bigg)\mathbf{K}_n+\begin{pmatrix}
        S_{\delta,b}^{(n)}\\
        S_{\Theta,b}^{(n)}\\
        S_{\delta,\chi}^{(n)}\\
        S_{\Theta,\chi}^{(n)}
    \end{pmatrix}\,,
\end{equation}
where the source terms for $n\geq 2$ are collected as
\begin{align}
    S_{\delta,I}^{(n)}&=\sum_{m=1}^{n-1}\frac{m!(n-m)!}{n!}\sum_{\substack{\cA\subset\{1,\dots,n\}\\ |\cA|=m}}\alpha(\bk_\cA,\bk_{\cA^c}) G_{m,I}(\{\bk_a\}_{a\in\cA})F_{n-m,I}(\{\bk_a\}_{a\in\cA^c})\,,\\
    S_{\Theta,I}^{(n)}&=\sum_{m=1}^{n-1}\frac{m!(n-m)!}{n!}\sum_{\substack{\cA\subset\{1,\dots,n\}\\ |\cA|=m}}\beta(\bk_\cA,\bk_{\cA^c}) G_{m,I}(\{\bk_a\}_{a\in\cA})G_{n-m,I}(\{\bk_a\}_{a\in\cA^c})\,.
\end{align}
with $\cA^c$ denoting the complement of $\cA$ in the set $\{1,\dots,n\}$ and $\bk_\cA=\sum_{a\in\cA}\bk_a$ such that the kernels are symmetrized.

To compute their contributions to the matter power spectrum, let us also introduce the total $m$ and relative $r$ contributions as
\begin{alignat*}{2}
    F_{n,m}&=w_bF_{n,b}+w_\chi F_{n,\chi}\,,\qquad && F_{n,r}=F_{n,\chi}-F_{n,b}\\
    G_{n,m}&=w_bG_{n,b}+w_\chi G_{n,\chi}\,,\qquad && G_{n,r}=G_{n,\chi}-G_{n,b}\,,
\end{alignat*}
where $w_\chi(N)=\rho_{\chi}(N)/(\rho_{b}(N)+\rho_{\chi}(N))$, $w_b=1-w_\chi$, and consequently $w_{\chi,N}=-c'\phi_{,N}w_bw_\chi$.
With these, the one-loop matter power spectrum can then be computed as
\begin{align}
    P_{22,m}(k,N)&=2\int \frac{\dd^3q}{(2\pi)^3}F_{2,m}(\bq,\bk-\bq;N)^2P_*(q)P_*(|\bk-bq|)\,,\\
    P_{13,m}(k,N)&=6F_{1,m}(k,N)P_*(k)\int\frac{\dd^3 q}{(2\pi)^3}F_{3,m}(\bk,\bq,-\bq;N)P_*(q)\,,
\end{align}
where $P_*$ is the power spectrum of the initial field $\delta_*$ we used to seed our perturbation kernels. 
At the observed time $N_{\rm obs}$, let $D_m(k)=F_{1,m}(k,N_{\rm obs})$ and define the linear total matter perturbation as $\delta_L(\bk)=D_m(k)\delta_*(\bk)$ such that the associated linear power spectrum satisfies
\begin{equation}
    \langle\delta_L(\bk)\delta_L(\bk')\rangle=(2\pi)^3\delta_D^3(\bk+\bk')P_L(k)\,,\qquad P_L(k)=D_m^2(k)P_*(k)\,.
\end{equation}
Furthermore, we introduce the reduced kernels as 
\begin{alignat*}{2}
    \bar{F}_{n,m}&=\frac{F_{n,m}}{\prod_{a=1}^n F_{1,m}(k_a,N)}\,,\qquad &&
    \bar{F}_{n,r}=\frac{F_{n,r}}{\prod_{a=1}^n F_{1,m}(k_a,N)}\,,\\
    \bar{G}_{n,m}&=\frac{G_{n,m}}{\prod_{a=1}^n F_{1,m}(k_a,N)}\,,\qquad &&
    \bar{G}_{n,r}=\frac{G_{n,r}}{\prod_{a=1}^n F_{1,m}(k_a,N)}\,,
\end{alignat*}
where we will use them for deducing the relevant kernels appearing in the galaxy power spectrum.

\begin{figure}
    \centering
    \begin{subfigure}{\textwidth}
    \begin{subfigure}{.49\textwidth}
    \includegraphics[width=\linewidth]{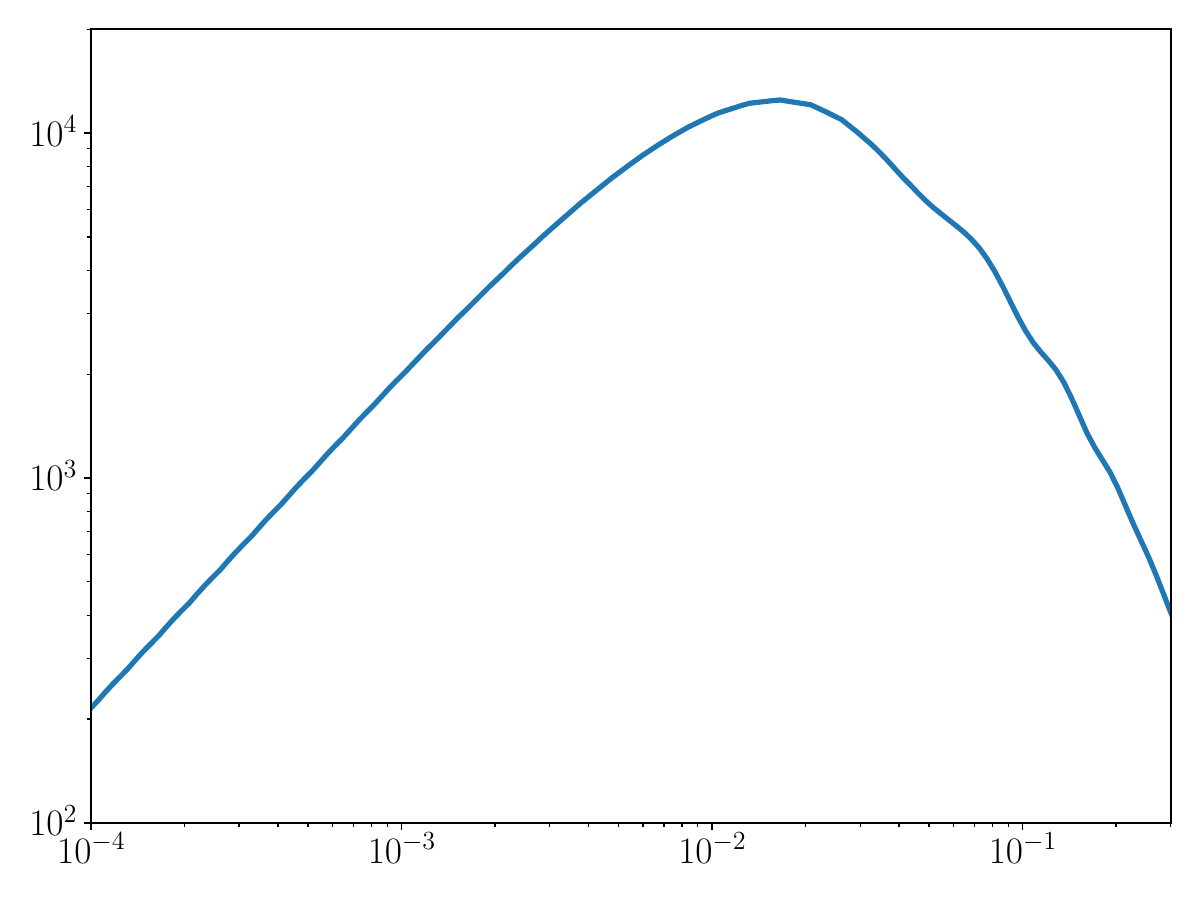}
    \begin{picture}(0,0)
        \put(-135,50){\rotatebox{90}{\footnotesize $P_{\rm tot}(k,z_{\rm LRG1})\,[(\mathrm{Mpc}/h)^3]$}}
        \put(-10, 0){\footnotesize $k\,[h\,\mathrm{Mpc}^{-1}]$}
    \end{picture}
    \end{subfigure}
    \hfill
    \begin{subfigure}{.49\textwidth}
    \includegraphics[width=\linewidth]{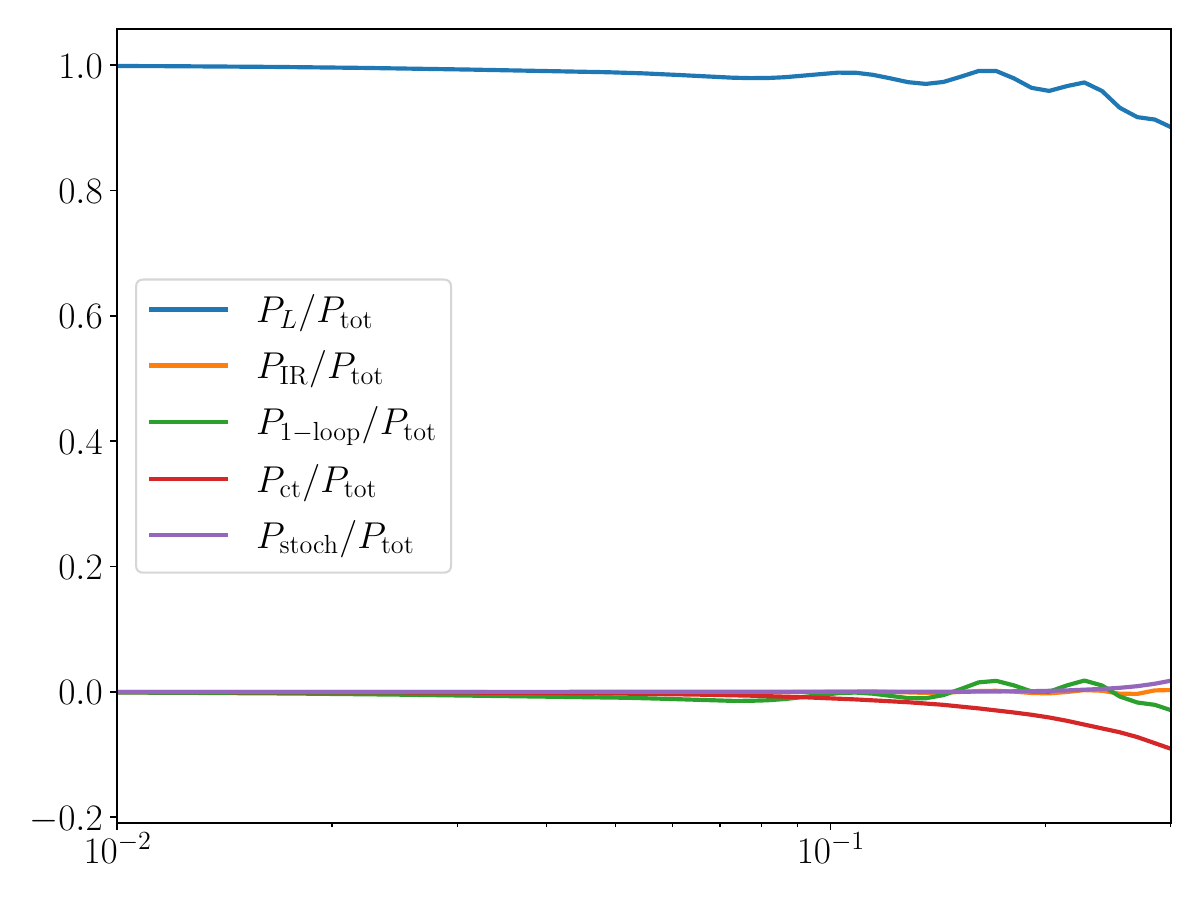}
    \begin{picture}(0,0)
        \put(-10, 0){\footnotesize $k\,[h\,\mathrm{Mpc}^{-1}]$}
    \end{picture}
    \end{subfigure}\vspace{-0.5cm}
    \caption{Matter power spectrum in real space at $z_{\rm LRG1}\approx 0.51$.}
    \end{subfigure}
    \begin{subfigure}{\textwidth}
    \begin{subfigure}{.49\textwidth}
    \includegraphics[width=\linewidth]{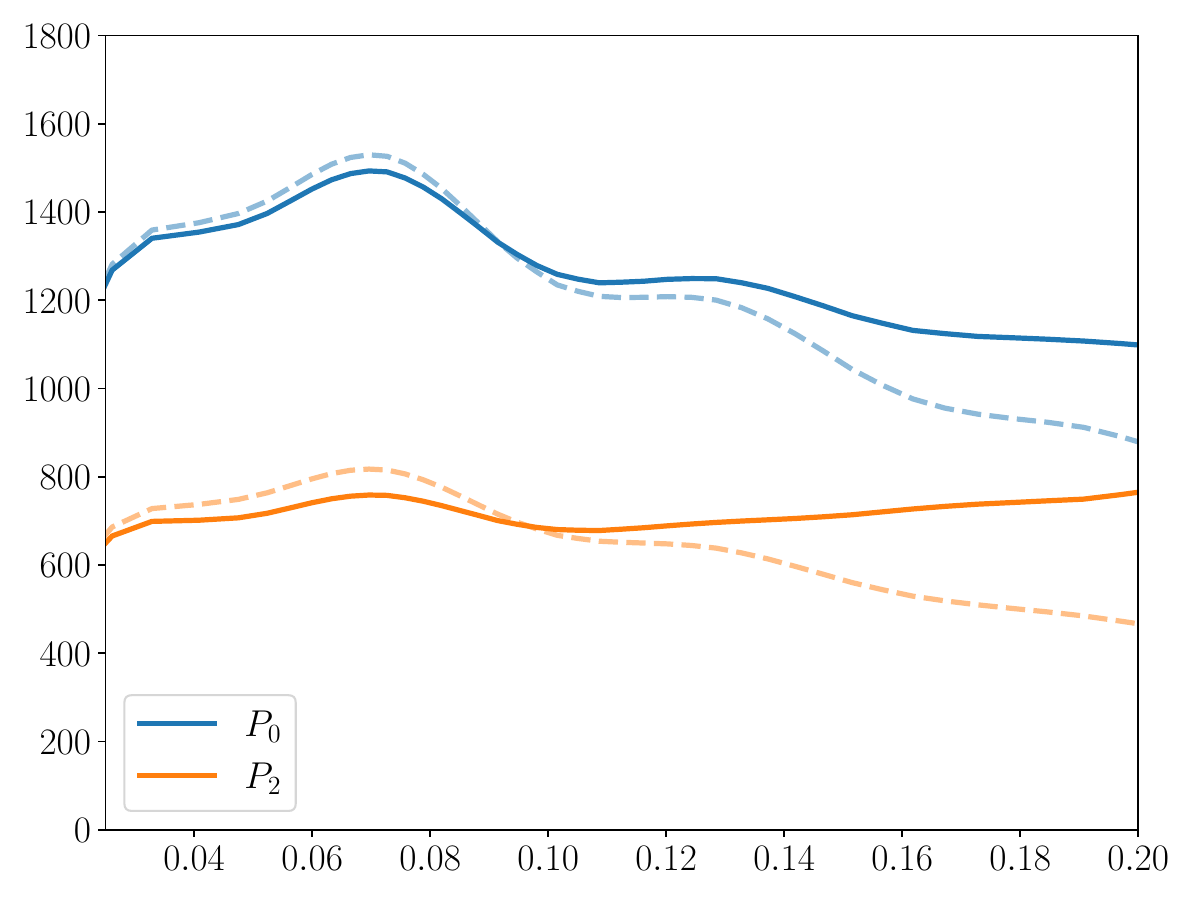}
    \begin{picture}(0,0)
        \put(-135,70){\rotatebox{90}{\footnotesize $kP_{l}(k)\,[(\mathrm{Mpc}/h)^2]$}}
        \put(-10, 0){\footnotesize $k\,[h\,\mathrm{Mpc}^{-1}]$}
    \end{picture}
    \end{subfigure}
    \hfill
    \begin{subfigure}{.49\textwidth}
    \includegraphics[width=\linewidth]{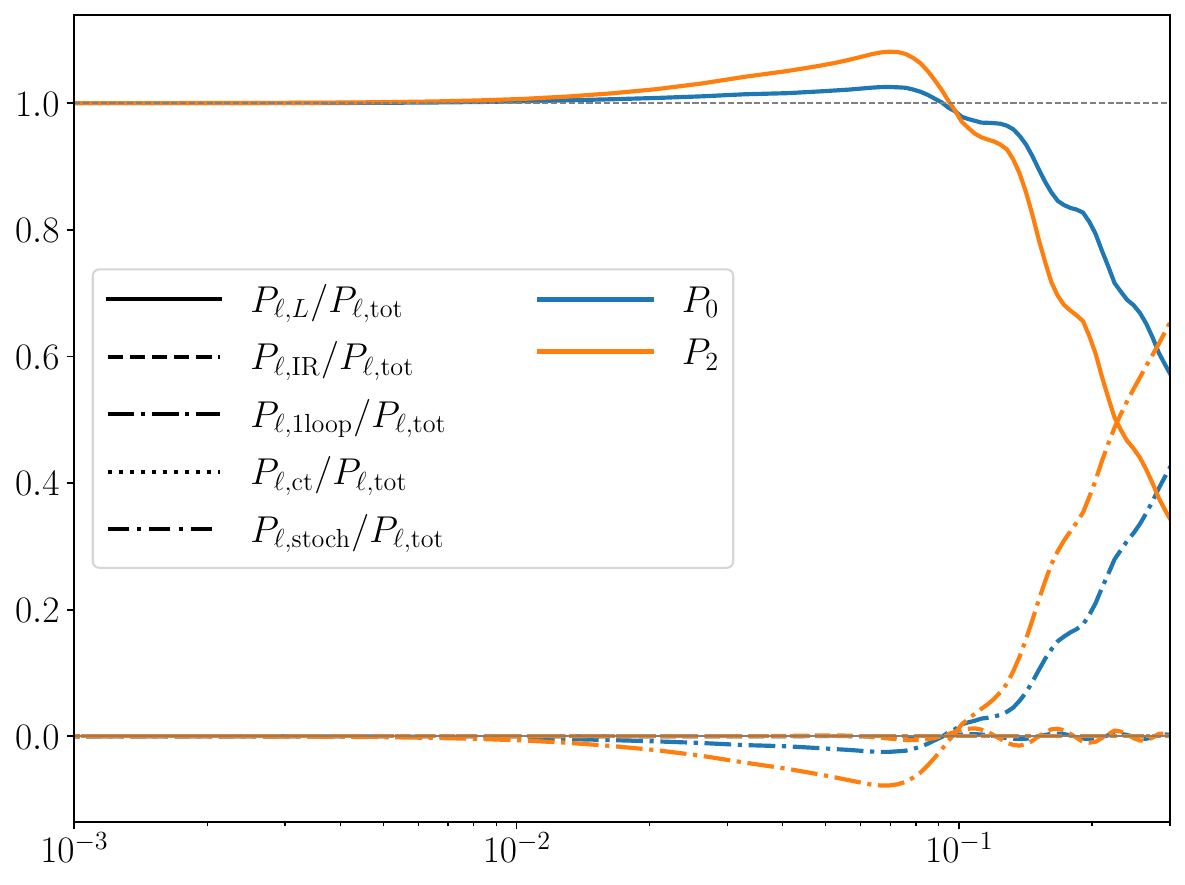}
    \begin{picture}(0,0)
        \put(-10, 0){\footnotesize $k\,[h\,\mathrm{Mpc}^{-1}]$}
    \end{picture}
    \end{subfigure}\vspace{-0.5cm}
    \caption{Galaxy power spectrum multipoles in redshift space}
    \end{subfigure}
    \caption{The best-fit power spectrum from CMB + DESI1.5 as an illustrative example of corrections to the matter power spectrum in the quasi-linear regime $k\sim 0.1\,h\,\mathrm{Mpc}^{-1}$. 
    }
    \label{fig:power spectrum breakdown}
\end{figure}

\subsection{The one-loop redshift-space galaxy power spectrum}

The galaxy overdensity is not determined solely by the linear matter density perturbation. In the effective field theory approach, the tracer overdensity is expanded in scalar operators constructed from the long-wavelength matter and velocity field, with coefficients absorbing unresolved astrophysical formation processes. We adopt the two-fluid Eulerian basis prescription carried out in~\cite{Schmidt:2016coo,Chen:2019cfu,Bottaro:2023wkd}
\begin{equation}
\label{eq:galaxy pert}
    \delta_g=b_m\delta_m+b_r\delta_r+\frac{b_{mm}}{2}[\delta_m^2]+b_{\cG mm}[s_m^2]+b_{\Gamma_3}[\Gamma_3]+\epsilon_g+\dots
\end{equation}
where $\delta_m=w_b\delta_b+w_\chi \delta_\chi,\delta_r=\delta_\chi-\delta_b$ denote the total and relative matter perturbations, respectively, $b$'s denote bias parameters in the theory~\cite{Desjacques:2016bnm,mcdonald2009clustering}, $[\cdot ]$ indicate renormalized bias operators, $\epsilon_g$ indicates the stochastic field, $\Gamma_3$ is associated with the difference between density and velocity tidal evolutions, and $s_{m,ij}=\big(\frac{\partial_i\partial_j}{\nabla^2}-\frac{\delta_{ij}}{3}\big)\delta_m$ denotes the trace-free tidal tensor.

Similar to the density fields, we can define the real-space tracer density kernels $D_n$ and velocity divergence kernels $V_n$ of galaxies by 
\begin{align}
\label{eq:realspace galaxy delta}
    \delta_g^{(n)}(\bk,N)&=\int\bigg[\prod_{a=1}^n\frac{\dd^3 k_a}{(2\pi)^3}\delta_L(\bk_a)\bigg](2\pi)^3\delta_D^{(3)}\bigg(\bk-\sum_{a=1}^n\bk_a\bigg)D_n(\bk_1,\dots,\bk_n,N)\,,\\
\label{eq:realspace galaxy theta}
    \Theta_g^{(n)}(\bk,N)&=\int\bigg[\prod_{a=1}^n\frac{\dd^3 k_a}{(2\pi)^3}\delta_L(\bk_a)\bigg](2\pi)^3\delta_D^{(3)}\bigg(\bk-\sum_{a=1}^n\bk_a\bigg)V_n(\bk_1,\dots,\bk_n,N)\,,
\end{align}
where $\delta_L(k)=F_{1,m}(k,N)\delta_*(k)$. We assume that galaxies co-move with the total matter fluid, $\mathbf{v}_g=\mathbf{v}_m=w_b\mathbf{v}_b+w_\chi \mathbf{v}_\chi$ and neglect independent relative velocity bias contributions. This assumption is further motivated by finding $(w_b G_{1,r}/G_{1,m})/(f\sigma_8)\sim 10^{-3}$ among the sampled points in our MCMC simulations, suggesting the correction to the redshift-space power spectrum is at the sub-percent level. With this assumption, we have
\begin{equation}
    V_n=\bar{G}_{n,m}\,,
\end{equation}
where no independent relative velocity galaxy bias parameters is sampled. Inserting $\delta_m$ into \cref{eq:galaxy pert}, we obtain
\begin{align}
    D_1&=b_m+b_r\bar{F}_{1,r}\,,\\ 
    D_2(\bk_1,\bk_2,N)&=b_m\bar{F}_{2,m}(\bk_1,\bk_2,N)+\frac{b_{mm}}{2}+b_{\cG mm}S_2(\bk_1,\bk_2,N)+b_r\bar{F}_{2,r}\,,\\
    D_3(\bk_1,\bk_2,\bk_3,N)&=b_m\bar{F}_{3,m}(\bk_1,\bk_2,\bk_3,N)+D_{3,\Gamma_3}(\bk_1,\bk_2,\bk_3,N)+b_r\bar{F}_{3,r}\nonumber\\
    &\qquad\qquad\qquad +\sum_{\substack{i\in\{1,2,3\}\\\{j,l\}\in\{1,2,3\}\backslash\{i\}}}\bigg[\frac{b_{mm}}{3}\bar{F}_{2,m}(\bk_j,\bk_l,N)+\frac{2b_{\cG mm}}{3}S_{i\vert jl}\bar{F}_{2,m}(\bk_j,\bk_l,N)\bigg]\,,
\end{align}
where $S_2(\bk_1,\bk_2)=\frac{(\bk_1\cdot\bk_2)^2}{k_1^2k_2^2}-\frac13$, $T_{i\vert jl}=\frac{(\bk_i\cdot(\bk_j+\bk_l))^2}{k_i^2|\bk_j+\bk_l|^2}-1$, $S_{i\vert jl}=T_{i\vert jl}+\frac23$, and
\begin{align}
    D_{3,\Gamma_3}(\bk_1,\bk_2,\bk_3,N)=\frac{2b_{\Gamma_3}}{3}\sum_{\substack{i\in\{1,2,3\}\\\{j,l\}\in\{1,2,3\}\backslash\{i\}}}T_{i\vert jl}\bigg[\bar{F}_{2,m}(\bk_j,\bk_l,N)-\frac{\bar{G}_{2,m}(\bk_j,\bk_l,N)}{f(N)}\bigg]\,,
\end{align}

Galaxies observed by DESI are measured via their angular positions and spectroscopic redshifts. The galaxy positions are hence shifted along the line-of-sight by their peculiar velocities. To compare with the full-shape data from DESI, we need to map the previous real-space kernels to the redshift-space. In the plane-parallel limit, the redshift-space position is~\cite{Ivanov:2018gjr,Lewandowski:2015ziq}
\begin{equation}
    \mathbf{s}=\mathbf{x}+u_{\parallel }(\mathbf{x})\hat{\mathbf{n}}\,,\qquad u_{\parallel}=\frac{\mathbf{v}_g\cdot\hat{n}}{\cH}\,,
\end{equation}
where $\mathbf{x}$ is the comoving real-space position, $\mathbf{v}_g=\dd \mathbf{x}/\dd \tau$, and $\tau$ is the conformal time. Conservation of galaxy number under this change of coordinates yields
\begin{equation}
    1+\delta^s_g(\mathbf{s})=\int \dd^3x[1+\delta_g(\mathbf{x})]\delta_D^{(3)}[\mathbf{s}-\mathbf{x}-u_{\parallel}(\mathbf{x})\hat{\mathbf{n}}]\,.
\end{equation}
Upon a Fourier transform and removing the homogeneous $\bk=0$ contribution yields 
\begin{equation}
\label{eq:redshift realspace}
    \delta_g^s(\bk)=\delta_g(\bk)+\int\dd^3xe^{-i\bk\cdot \mathbf{x}}(1+\delta_g(\mathbf{x}))\big(e^{-ik_\parallel u_\parallel(\mathbf{x})}-1\big)\,,
\end{equation}
with $k_\parallel=\bk\cdot\hat{\mathbf{n}}=k\mu$ and $\mu=\hat{k}\cdot \hat{n}$. Assuming an irrotational large-scale velocity $\mathbf{v}_g(\bk)=i\cH \bk\Theta_g(\bk)/k^2$, we have $u_\parallel^{(n)}(\bk)=i\frac{\bk\cdot \hat{\mathbf{n}}}{k^2}\Theta_g^{(n)}(\bk)$. The redshift-space kernels $Z_n$ can then be defined by
\begin{equation}
\label{eq:redshift kernel}
    \delta_g^{s,(n)}(\bk)=\int\bigg[\prod_{a=1}^n\frac{\dd^3 k_a}{(2\pi)^3}\delta_L(\bk_a)\bigg](2\pi)^3\delta_D^{(3)}\bigg(\bk-\sum_{a=1}^n\bk_a\bigg)Z_n(\bk_1,\dots,\bk_n;\hat{\mathbf{n}})\,. 
\end{equation}
Then, plugging \eqref{eq:redshift kernel}, \eqref{eq:realspace galaxy delta} and chugging \eqref{eq:realspace galaxy theta} into \cref{eq:redshift realspace} while expanding the exponential, we obtain
\begin{align}
    Z_1(\bk,\mu,N)&=D_1(\bk,N)+\mu^2V_1(\bk,N)\,,\\
    Z_2(\bk_1,\bk_2,N)&=D_2(\bk_1,\bk_2,N)+\mu^2V_2(\bk_1,\bk_2,N)\nonumber\\
    &\qquad\qquad\qquad\qquad+\frac{k_\parallel}{2}[U_1(\bk_1,N)Z_1(\bk_2,\mu_2,N)+U_1(\bk_2,N)Z_1(\bk_1,\mu_1,N)]\,,\\
    Z_3(\bk_1,\bk_2,\bk_3,N)&=D_3(\bk_1,\bk_2,\bk_3,N)+\mu^2V_3(\bk_1,\bk_2,\bk_3,N)\nonumber\\
    &\qquad\qquad\qquad\qquad+\sum_{\substack{i\in\{1,2,3\}\\\{j,l\}\in\{1,2,3\}\backslash \{i\}}}\frac{k_\parallel}{3}U_1(\bk_i,N)[D_2(\bk_j,\bk_l,N)+\mu_{jl}^2V_2(\bk_j,\bk_l,N)]\nonumber\\
    &\qquad\qquad\qquad\qquad+\frac{k_\parallel}{3}U_2(\bk_j,\bk_l,N)Z_1(\bk_i,\mu_i,N)+\frac{k_\parallel^2}{6}U_1(\bk_j,N)U_1(\bk_l,N)Z_1(\bk_i,\mu_i,N)\,,
\end{align}
where we have used
\begin{equation}
    U_1(\bk_i,N)=\frac{\mu_i}{k_i}V_1(\bk_i,N)\,,\qquad U_2(\bk_j,\bk_l,N)=\frac{\mu_{jl}}{|\bk_j+\bk_l|}V_2(\bk_j,\bk_l,N)\,,\qquad \mu_{jl}=\frac{(\bk_j+\bk_l)\cdot{\hat{n}}}{|\bk_j+\bk_l|}\,.
\end{equation}

The above one-loop convention removes the independent local quadratic-bias contribution from $P_{13}$. We can see this simplification by introducing the standard local quadratic coordinate $\tilde{b}_2=b_{mm}+\frac43 b_{\cG mm}$. The projection acts on the third-order terms
\begin{align}
    D_2^{\rm ren}&=D_2-\frac12 \tilde{b}_2\,,\\
    D_3^{\rm ren}&=D_3-\frac{\tilde{b}_2}{3}\sum_{\substack{i\in\{1,2,3\}\\\{j,l\}\in\{1,2,3\}\backslash \{i\}}} \bar{F}_{2,m}(\bk_j,\bk_l)\,,
\end{align}
where $D_2$ enters via $U_1D_2$. Therefore, the third-order redshift-space kernel contributing to $P_{13}$ is
\begin{multline}
    Z_3^{\rm ren}=D_3^{\rm ren}+\mu^2\bar{G}_{3,m}+\sum_{\substack{i\in\{1,2,3\}\\\{j,l\}\in\{1,2,3\}\backslash \{i\}}}\frac{k_\parallel}{3}U_1(\bk_i)[D_2^{\rm ren}(\bk_j,\bk_l)+\mu_{jl}^2\bar{G}_{2,m}(\bk_j,\bk_l)]\\
    +\frac{k_\parallel}{3}U_2(\bk_j,\bk_l)Z_1(k_i,\mu_i)+\frac{k_\parallel^2}{6}U_1(\bk_j)U_1(\bk_l)Z_1(k_i,\mu_i)\,,
\end{multline}
where we have suppressed the time dependence.

From here, we can construct the perturbative contributions to the galaxy power spectrum
\begin{align}
    P_{\rm tree}(k,\mu)&=Z_1^2(k,\mu)P_L(k)\,,\\
    P_{22}(k,\mu)&=2\int \frac{\dd^3q}{(2\pi)^3}Z_2^2(\bq,\bp)P_L(q)P_L(p)\,,\\
    P_{13}(k,\mu)&=6Z_1(k,\mu)P_L(k)\int\frac{\dd^3q}{(2\pi)^3}Z_3^{\rm ren}(\bk,\bq,-\bq)P_L(q)\,,
\end{align}
where $\bp=\bk-\bq$.

\subsection{Counterterms and stochastic contributions}

In addition to the one-loop contributions, the power spectrum also include analytic counterterms for unresolved short-distance dynamics and a stochastic contribution associated with discrete tracers
\begin{align}
\label{eq:counterterm}
    P_{\rm ct}(k,\mu)&=-k^2P_L(k)(c_0+c_2\mu^2+c_4\mu^4)\,,\\
\label{eq:stochastic}
    P_{\rm stoch}(k,\mu)&=N_0+k^2(N_2+N_{2\mu}\mu^2)\,,
\end{align}
where $c_0,c_2,c_4$ are the redshift-space EFT coefficients while $N_0,N_2,N_{2\mu}$ are the native stochastic amplitudes.
Given a galaxy sample, we analytically marginalize four Gaussian-prior variables $\alpha_0,\alpha_2,\text{SN}_0,\text{SN}_2$, as was done in~\cite{DESI:2024hhd}, with their mapping to the nuisance parameters as
\begin{alignat*}{3}
    c_0&=-\frac{b_m^2}{h^2}\alpha_0\,,\qquad && c_2=-\frac{b_mf_{\rm eff}}{h^2}(\alpha_0+\alpha_2)\,,\qquad &&c_4=-\frac{f_{\rm eff}^2}{h^2}\alpha_2\,,\\
    N_0&=\frac{P_{\rm shot}}{h^3}\text{SN}_0\,,\qquad &&N_2=0\,,\qquad &&N_{2\mu}=\frac{P_{\rm shot}f_{\rm sat}\sigma_v^2}{h^5}\text{SN}_2\,,
\end{alignat*}
where $f_{\rm eff}$ is the effective linear growth rate, and $P_{\rm shot},f_{\rm sat},\sigma_v$ are tracer-specific shot-noise normalizations, satellite fraction, and stochastic velocity scale used in the likelihood.

\subsection{Anisotropic IR resummation}

Long-wavelength displacements can smear the BAO features of the galaxy power spectrum. To resum these motions, we can split the linear matter power spectrum into a smooth component $P_{\rm nw}$ and an oscillatory component $P_w$,
\begin{equation}
    P_L(k)=P_{\rm nw}(k)+P_w(k)\,.
\end{equation}
Then, the displacement variances associated to the density $D$ and line-of-sight $V$ displacement responses are
\begin{align}
    \Sigma_{DD}&=\frac{1}{6\pi^2}\int_{q_{\rm min}}^{k_S} \dd q P_{\rm nw}(q)J(qr_{\rm BAO})\,,\\
    \Sigma_{DV}&=\frac{1}{6\pi^2}\int_{q_{\rm min}}^{k_S} \dd q P_{\rm nw}(q)g(q)J(qr_{\rm BAO})\,,\\
    \Sigma_{VV}&=\frac{1}{6\pi^2}\int_{q_{\rm min}}^{k_S}\dd q P_{\rm nw}(q)g^2(q)J(qr_{\rm BAO})\,,\\
    \Delta\Sigma_{VV}&=\frac{1}{2\pi^2}\int_{q_{\rm min}}^{k_S}\dd q P_{\rm nw}(q)g^2(q)j_2(qr_{\rm BAO})\,,
\end{align}
where $r_{\rm BAO}$ is the characteristic BAO separation, $k_S$ is the IR/UV separation wavenumber, $q_{\rm min}$ the numerical lower limit of the displacement integration, $j_l(x)$ the spherical Bessel function of order $l$, $J(x)=1-j_0(x)+2j_2(x)$, and $g(q)=\bar{G}_{1,m}(q)$. The anisotropic damping exponent is 
\begin{equation}
    S(k,\mu)=k^2\Sigma_s^2(\mu)\,,
\end{equation}
with $\Sigma_s^2(\mu)=\Sigma_{DD}+2\mu^2\Sigma_{DV}+\mu^2\Sigma_{VV}+\mu^2(\mu^2-1)\Delta\Sigma_{VV}$. In the scale-independent case, we can introduce a constant $f=g(q)$ such that our variances can be identified with $\Sigma^2,\delta\Sigma^2$ introduced in~\cite{Ivanov:2018gjr} where $\Sigma_{DD}=\Sigma^2, \Sigma_{DV}=f\Sigma^2,\Sigma_{VV}=f^2\Sigma^2,\Delta\Sigma_{VV}=f^2\delta\Sigma^2$, reproducing its known behavior in this limit.

At one loop, the IR prescription associated to one wiggle in the power spectrum is~\cite{Blas:2016sfa}
\begin{align}
    P^{\rm IR}_{\rm tree}&=Z_1^2(k,\mu)[P_{\rm nw}(k)+e^{-S}(1+S)P_w(k)]\,,\\
    P^{\rm IR}_{\rm 1-loop}(k,\mu)&=P_{\rm 1-loop,nw}(k,\mu)+e^{-S}P_{\rm 1-loop,w}(k,\mu)\,,
\end{align}
where $P_{\rm 1-loop}=P_{22}+P_{13}$. Thus, $P_{22}$ includes two permutations of one smooth and one wiggly power spectrum while $P_{13}$ includes wiggle insertions in its external or internal linear power factor. The EFT counterterm uses the damped combination $P_{\rm nw}+e^{-S}P_w$ while the stochastic term is unaffected. Hence, the final galaxy power spectrum is
\begin{multline}
\label{eq:final power spectrum}
    P_g^{s,{\rm IR}}(k,\mu)=Z_1^2[P_{\rm nw}+e^{-S}(1+S)P_w]+P_{\rm 1-loop,nw}+e^{-S}P_{\rm 1-loop,w}\\
    -k^2 P_{L,{\rm ct}}^{\rm IR}(c_0+c_2\mu^2+c_4\mu^4)+N_0+k^2(N_2+N_{2\mu}\mu^2)\,.
\end{multline}

\begin{table}[tp]
    \centering
    \begin{tabular}{llll}
        \hline\hline
        Type & Parameter & Prior & Units \\
        \hline

        \textbf{nuisance}
        & $b_1^{E}\sigma_8(z)$
        & $\mathcal{U}[0,3]$
        & --- \\

        \textbf{(sampled)}
        & $b_2^{L}\sigma_8^2(z)$
        & $\mathcal{N}(0,5^2)$
        & --- \\

        &
        $b_s^{L}\sigma_8^2(z)$
        & $\mathcal{N}(0,5^2)$
        & --- \\

        \hline

        \textbf{nuisance}
        & $\alpha_0$
        & $\mathcal{N}(0,12.5^2)$
        & $[{\rm Mpc}/h]^2$ \\

        \textbf{(analytically}
        & $\alpha_2$
        & $\mathcal{N}(0,12.5^2)$
        & $[{\rm Mpc}/h]^2$ \\

        \textbf{marginalized)}
        & ${\rm SN}_0$
        & $\mathcal{N}(0,2^2)$
        & --- \\

        &
        ${\rm SN}_2$
        & $\mathcal{N}(0,5^2)$
        & --- \\

        \hline
    \end{tabular}
    \caption{The nuisance parameters and priors used for each DESI full-shape tracer. Here, $\mathcal{U}[a,b]$ denotes a uniform prior between $a$ and $b$,while $\mathcal{N}(\mu,\sigma^2)$ denotes a Gaussian prior with mean $\mu$ and variance $\sigma^2$. The bias combinations $b_1^{E}\sigma_8(z)$,$b_2^{L}\sigma_8^2(z)$, and $b_s^{L}\sigma_8^2(z)$ are sampled explicitly in the MCMC chains. The counterterm parameters $\alpha_0$ and $\alpha_2$ and stochastic amplitudes ${\rm SN}_0$ and ${\rm SN}_2$ enter linearly and are analytically marginalized in the likelihood. The same nuisance prescription is applied independently to each full-shape redshift/tracer bin.}
    \label{tab:nuisance_priors}
\end{table}

\subsection{Alcock--Paczynski mapping}

To compare a theory with different expansion history to measurements in the coordinates adapted by observations, we apply the Alcock--Paczynski (AP) transformation~\cite{Alcock:1979mp}. At redshift $z$, we adapt the notation used in~\cite{eBOSS:2022pea}
\begin{equation}
    \alpha_\perp =\frac{D_M(z)}{D_{M,{\rm fid}}(z)}\,,\qquad \alpha_\parallel=\frac{D_H(z)}{D_{H,{\rm fid}}(z)}\,,
\end{equation}
where $D_M(z)$ is the transverse comoving distance and the subscript $fid$ refers to the fiducial cosmology used in constructing the survey coordinates. Let $k_f,\mu_f$ be the wavenumber and line-of-sight angle in the fiducial cosmology, then the theory coordinates and volume Jacobian are~\cite{Chudaykin:2020aoj}
\begin{align}
    k_t&=\frac{k_f}{\alpha_\perp}\sqrt{1+\mu_f^2\bigg(\frac{\alpha_\perp^2}{\alpha_\parallel^2}-1\bigg)}\,,\\
    \mu_t&=\frac{\mu_f\alpha_\perp/\alpha_\parallel}{\sqrt{1+\mu_f^2(\alpha_\perp^2/\alpha_\parallel^2-1)}}\,,\\
\label{eq:fiducial mapping}
    P_{g,{\rm fid}}^{s}(k_f,\mu_f)&=\frac{P_g^{s,{\rm IR}}(k_t,\mu_t)}{\alpha_\perp^2\alpha_\parallel}\,.
\end{align}
Using this last fiducial power spectrum, the observed-coordinate multipoles are 
\begin{equation}
    P_{l}^{\rm fid}(k_f)=\frac{2l+1}{2}\int_{-1}^1 \dd\mu_f \cL_l(\mu_f)P_{g,{\rm fid}}^s(k_f,\mu_f)\,,
\end{equation}
where $l=0,2,4$ and $\cL_l$ is the Legendre polynomial.

The behavior of the matter power spectrum along with the multipole moments of the redshift-space galaxy power spectrum for the FDS model is shown in \cref{fig:power spectrum breakdown}.

\begin{figure}[!tp]
    \centering
    \includegraphics[width=\linewidth]{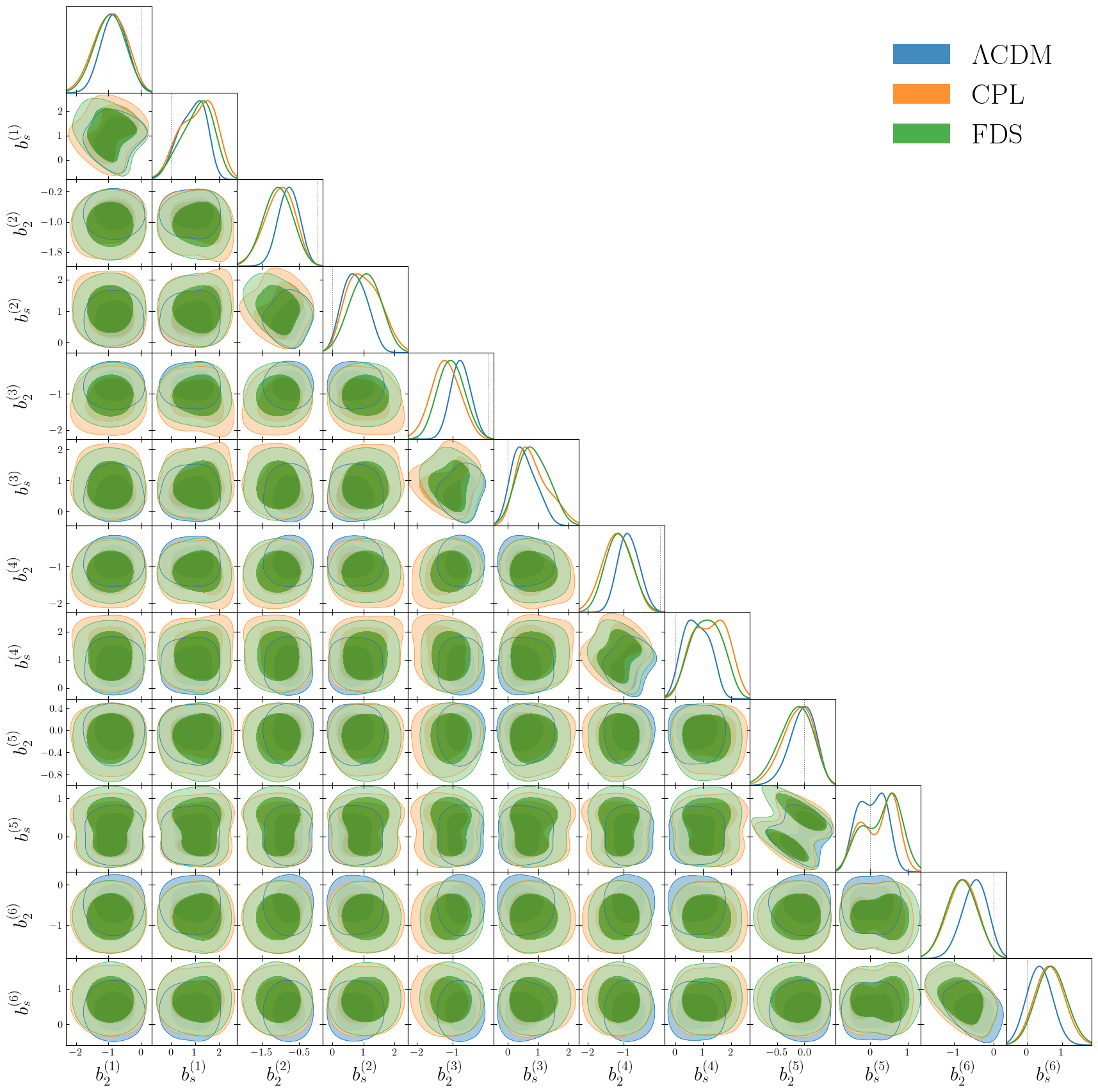}
    \caption{Comparison of the quadratic galaxy bias parameters among $\Lambda$CDM, the CPL parameterization, and the FDS model where the cosmology is fitted to CMB+DESI1.5. Here, $b_2^{(i)}$ and $b_{s}^{(i)}$ correspond to the quadratic local and tidal Lagrangian bias parameters. The superscript $i$ runs from 1 to 6 indicating the galaxy bias parameter for BGS, LRG1, LRG2, LRG3, ELG2, and QSO, respectively, in the DESI DR1 database.}
    \label{fig:bias tracers}
\end{figure}

\section{Mapping to DESI multipoles}
\label{sec:DESI multipoles}

The galaxy-bias expansion \eqref{eq:realspace galaxy delta} was carried out using the Eulerian coefficients $b_i$'s. However, the bias parameters relevant to the DESI full-shape analysis is done with Lagrangian bias coefficients. In Lagrangian perturbation theory, galaxies associated with an initial position $\bq$ has the following expansion
\begin{equation}
    \delta_g^L(\bq)=b_1^L\delta_L(\bq)+\frac{b_2^L}{2}[\delta_L^2]+b_s^L[s_L^2(\bq)]+\dots\,,
\end{equation}
where this is related to the Eulerian positions as $\mathbf{x}=\bq+\Psi(\bq)$. For a conserved, co-moving tracer population, galaxy number conservation implies
\begin{equation}
    1+\delta_g^E(\mathbf{x})=[1+\delta_m^E(\mathbf{x})][1+\delta_g^L(\bq)]\,.
\end{equation}
Then, the mapping between the Eulerian and Lagrangian bases at one-loop order becomes~\cite{Desjacques:2016bnm,Chen:2020fxs}
\begin{align}
    b_m^E&=1+b_1^L\,,\\
    b_{mm}^E&=b_2^L+\frac{8}{21}b_1^L\,,\\
    b_{\cG mm}^E&=b_s^L-\frac27 b_1^L\,.
\end{align}
In this basis, the DESI nuisance parameters then appear explicitly.
The simulation results for the sampled quadratic bias parameters are shown in figure~\ref{fig:bias tracers}.

Additionally, to make contact with observed galaxy surveys, we need to select an angular mask and redshift that is measured. Let $a$ index an observed power-spectrum multipole and wavenumber bin, and $b$ index the ordered pre-window theory entries. Let $a=(l,i),b=(l',j)$ denote the multipole moments $l,l'$ and the wavenumber bins $i,j$, then the window matrix $W_{ab}$ maps the fiducial coordinate theory into the observed data space via
\begin{equation}
\label{eq:DESI mapping}
    m_a(\mathbf{\vartheta},\mathbf{\eta})=\sum_{b}W_{ab}P_{b}^{\rm fid}(\mathbf{\vartheta},\mathbf{\eta})\,,
\end{equation}
where $\mathbf{\vartheta}$ denotes cosmological parameters and $\mathbf{\eta}$ denotes the galaxy-bias, counterterm, and stochastic nuisance parameters. The actual window differs between observations. The relevant DESI window matrix can be found in~\cite{DESI:2024aax}. 
Hence, combining \cref{eq:final power spectrum,eq:fiducial mapping,eq:DESI mapping}, we can specify the observable one-loop redshift-space multipoles adequate for fitting to the DESI data appropriate for the fading dark sector.

\end{document}